\RequirePackage{amsmath}
\documentclass[pre,amsmath,amssymb,groupedaddress,graphicx,showkeys,showpacs,reprint]{revtex4-1}

\usepackage[colorlinks=true,linkcolor=black,citecolor=black,bookmarks]{hyperref}
\usepackage{url}
\usepackage{units}
\usepackage{graphicx}
\usepackage{subfigure}
\usepackage[utf8]{inputenc}
\usepackage{ae}
\usepackage{natbib}
\usepackage{multirow}

\usepackage{bm} 
\usepackage[group-separator={,}]{siunitx}

\renewcommand{\vec}[1]{\boldsymbol{#1}}
\newcommand\eg{e.g.}
\newcommand\ie{i.e.}
\newcommand{\dc}[1]{\varepsilon_{#1}}

\begin{document}

\title{Characterization of Maximally Random Jammed Sphere Packings.\\ III. Transport and Electromagnetic Properties via Correlation Functions}

\author{Michael A. Klatt}
   \affiliation{Karlsruhe Institute of Technology (KIT), Institute for Stochastics, Englerstr. 2, 76131 Karlsruhe, Germany}
\author{Salvatore Torquato}
   \email[Electronic mail: ]{torquato@electron.princeton.edu}
   \affiliation{Department of Chemistry, Department of Physics, Princeton
     Institute for the Science and Technology of Materials, and Program
     in Applied and Computational Mathematics, Princeton University, Princeton, New Jersey 08544, USA}
\date{\today}

\begin{abstract}
  In the first two papers of this series, we characterized the structure of maximally random jammed (MRJ) sphere packings across length scales by computing a variety of different correlation functions, spectral functions, hole probabilities, and local density fluctuations.
  From the remarkable structural features of the MRJ packings, especially its disordered hyperuniformity, exceptional physical properties can be expected.
  Here, we employ these structural descriptors to estimate effective transport and electromagnetic properties
  via rigorous bounds, exact expansions, and accurate analytical approximation formulas.
  These property formulas include interfacial bounds as well as universal scaling laws for the mean survival time and the fluid permeability.
  We also estimate the principal relaxation time associated with Brownian motion among perfectly absorbing ``traps.''
  For the propagation of electromagnetic waves in the long-wavelength limit, we show that a dispersion of dielectric MRJ spheres within a matrix of another dielectric material forms,
  to a very good approximation, a dissipationless disordered and isotropic two-phase medium for any phase dielectric contrast ratio.
  We compare the effective properties of the MRJ sphere packings to those of overlapping spheres, equilibrium hard-sphere packings, and lattices of hard spheres.
  Moreover, we generalize results to micro- and macroscopically anisotropic packings of spheroids with tensorial effective properties.
  The analytic bounds predict the qualitative trend in the physical properties associated with these structures, which provides guidance to more time-consuming simulations and experiments.
  They especially provide impetus for experiments to design materials with unique bulk properties resulting from hyperuniformity, including structural-color and color-sensing applications.
\end{abstract}

\pacs{}{}
\maketitle


\section{Introduction}
\label{Introduction}

Using results from homogenization theory that links statistical correlation functions to effective properties, we estimate effective transport and electromagnetic properties of maximally random jammed (MRJ) packings of identical spheres distributed throughout a matrix (or void) phase.
Packings of hard spheres exhibit a rich multitude of states~\cite{Henley1986, Finney1997, Zallen1998, bertrand_protocol_2016} and serve as simple, yet effective models of diverse many-particle systems or heterogeneous materials~\cite{ChaikinLubensky2000, ManoharanEtAl2003, Zohdi2006, HansenMcDonald2006, Torquato2002}.
An especially remarkable structure among the set of all isotropic, frictionless and statistically homogeneous sphere packings has been found in the MRJ state ~\cite{TorquatoEtAl2000PhysRevLetRCPvsMRJ}.
Intuitively speaking, MRJ packings are the maximally disordered among all mechanically stable packings.
More precisely, they minimize among the jammed packings some order metric $\Psi$~\cite{TorquatoEtAl2000PhysRevLetRCPvsMRJ, TorquatoStillinger2010RevModPhys, ohern_PRL_2002, Karayiannis_PRL_2008, XuRice2011, Ozawa_etal_2012, Baranau_etal_2013, tian_geometric-structure_2015, ramola_disordered_2016}.
The MRJ state can be unambiguously identified for a particular choice of the order metric, and
a variety of sensible, positively correlated order metrics produce an MRJ state in three dimensions with the same packing fraction 0.64~\cite{TorquatoStillinger2010RevModPhys}.
This definition is mathematically more precise than the familiar notion of random close packing (RCP)~\cite{Bernal1960, ScottKilgour1969, Finney1970, Berryman1983, HernEtAl2003, Aste2005, KuritaWeeks2011, Berthier_etal_2011, KapferEtAl2012}.
Moreover, while similar packing fractions have been reported for three-dimensional RCP and MRJ packings of identical spheres~\cite{TorquatoEtAl2000PhysRevLetRCPvsMRJ, HernEtAl2003, ScottKilgour1969},
there are other structural attributes that differ distinctly~\cite{TorquatoStillinger2010RevModPhys, Ji11, AtkinsonEtAl2013}.
The distinction between these states become especially vivid in two dimensions \cite{At14}.

The most remarkable structural feature of MRJ packings is their singular property of hyperuniformity~\cite{TorquatoStillinger2003, ZacharyTorquato2009}.
There is an anomalous suppression of infinite-wavelength density (volume-fraction) fluctuations, which results in $d$-dimensional Euclidean space $\mathbb{R}^d$ in negative quasi-long-range pair correlations that decay asymptotically like $-1/r^{d+1}$~\cite{DonevEtAl2005, ZacharyJiaoTorquato2011}.
The disordered hyperuniformity of the MRJ state can be interpreted as an `inverted critical phenomenon'
because the direct correlation function becomes long-ranged, in contrast to thermal critical points in which this function remains short-ranged~\cite{TorquatoStillinger2003, HopkinsStillingerTorquato2012}.

In this series of papers, we characterize in detail the complex and subtle structure and physical properties of three-dimensional MRJ packings of identical spheres.
In the first paper~\cite{KlattTorquato2014}, we structurally characterized MRJ sphere packings generated in Ref.~\cite{AtkinsonEtAl2013} using Voronoi statistics, including certain types of correlation functions.
We compared these computations to corresponding calculations for overlapping spheres and equilibrium hard spheres.

In the second paper~\cite{KlattTorquato2016}, we further characterized the MRJ sphere packings using the void-void, surface-void, and surface-surface
correlation functions, as well as, the spectral density, the exclusion (``hole'') probability, and density fluctuations.

In this paper, we use these structural characteristics to predict a variety of effective properties of the MRJ packings via rigorous bounds,
exact expansions, and accurate analytical approximation formulas.
Such a characterization of the microstructure provides similar bounds on seemingly unrelated physical effective properties,
and it allows one to apply cross-property relations~\cite{Torquato1990, AvellanedaTorquato1991}.
Here we specifically estimate the mean survival time~\cite{Prager1963, TorquatoYeong1997} and principal relaxation time associated with Brownian motion
of point particles among spherical traps~\cite{TorquatoAvellaneda1991, Torquato2002}, fluid permeability~\cite{Prager1961,BerrymanMilton1985, RubinsteinTorquato1989JFM}, and the effective complex dielectric constant associated with the propagation of electromagnetic waves\cite{RechtsmanTorquato2008}.

We compare the property predictions for MRJ packings to those for overlapping spheres (ideal gas) and equilibrium hard spheres.
The analytic bounds allow one to compare the properties of these different complex microstructures without time-expensive computer simulations.
It turns out that the property bounds capture well the qualitative trend of the actual physical properties, which is confirmed by universal scaling laws.

Most importantly, we study the effect of \textit{disordered hyperuniformity} on the propagation of electromagnetic waves.
Because of the anomalous suppression of large-scale density fluctuations, a hyperuniform material is nearly dissipationless in the long-wavelength limit~\cite{RechtsmanTorquato2008}.
This attribute could be useful for the design of photonic materials with novel structural color characteristics ~\cite{Ballato:00,ADMA:ADMA200903693,PhysRevE.90.062302,wilts_butterfly_2017} or color-sensing capabilities~\cite{PhysRevE.89.022721}.
It was shown that the imaginary part of the effective dielectric constant vanishes exactly through lowest order in perturbation expansion and third order in the wave vector.
Here we apply these results to demonstrate how the imaginary part of the MRJ sphere packings is orders of magnitude smaller than those of the overlapping or equilibrium hard spheres.

The rest of the paper is organized as follows.
In Sec.~\ref{sec:permeability-trapping}, we determine both void and interfacial bounds for the fluid permeability and the mean survival time.
The analytic calculations of the corresponding integrals over the correlation functions for finite packings and the extrapolation to infinite system sizes are summarized in Appendix~\ref{sec_integration_S2}.
In Appendix~\ref{sec_lattices}, the pore-size bounds on the mean survival time are calculated for lattices.
In Sec.~\ref{sec:mean-survival-time}, we consider the principal diffusion relaxation.
The effective dielectric constant is studied in Sec.~\ref{sec:eff_dielec}.
Finally, in Sec.~\ref{sec_spheroids}, tensorial effective physical properties are evaluated for anisotropic packing of oriented spheroids.
We summarize the results and make concluding remarks in Sec.~\ref{sec:Conclusion}.


\section{Fluid permeability and mean survival time}
\label{sec:permeability-trapping}

\begin{figure}[t]
  \centering
  \includegraphics[width=\linewidth]{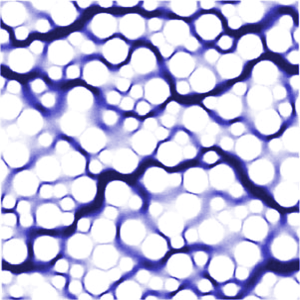}
  \caption{(Color online) Laminar flow through the void space of a two-dimensional hard-disk packing.
  The color (gray value) indicates the magnitude of the flow velocity.
  The image is reproduced from Ref.~\cite{Torquato2002} by permission from Springer.}
  \label{fig:visualization-flow}
\end{figure}

The fluid permeability $k$ characterizes slow laminar flow of an incompressible viscous fluid through porous media according to Darcy's law~\cite{DarcyFootnote}
\begin{align}
  \mathbf{U} = -\frac{k}{\mu} \nabla p_0,
\end{align}
where $\mathbf{U}$ is the average fluid velocity, $\nabla p_0$ is the applied pressure gradient, and $\mu$ is the dynamic viscosity~\footnote{For more details on the definition of these quantities and the notation, see Ref.~\cite[][Sec.~13.5]{Torquato2002}.}.
Figure~\ref{fig:visualization-flow} shows an example of a two-dimensional flow through the void space of a two-dimensional hard-disk packing. 

\begin{figure}[t]
  \centering
  \includegraphics[width=\linewidth]{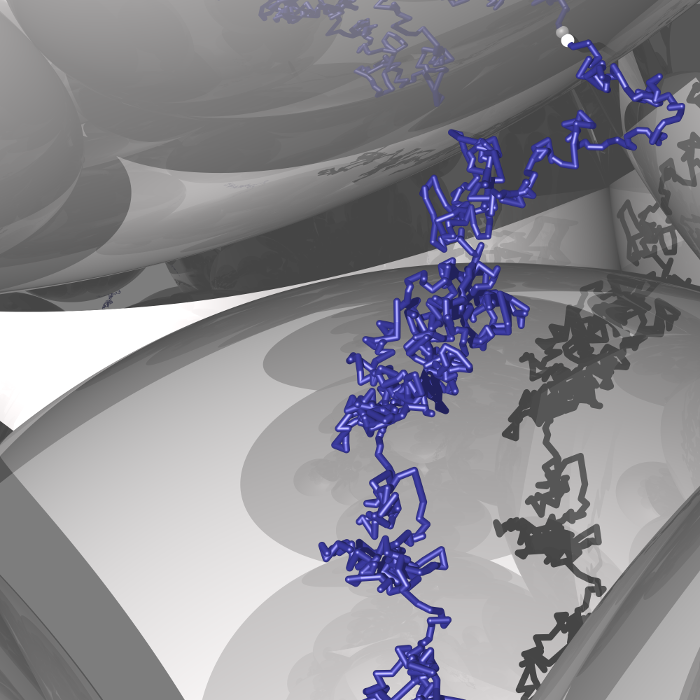}
  \caption{(Color online) Diffusion of a particle through the void space of a three-dimensional packing of perfectly absorbing spheres.
    The blue trajectory shows an approximation of its Brownian motion.
  The particle is absorbed once it touches the surface of a sphere (indicated by a small white ball).}
  \label{fig:visualization-diffusion}
\end{figure}

The mean survival time $\tau$ and the trapping constant~$\gamma$ characterize diffusion and reaction among absorbing ``traps'' that are static~\footnote{For more details, on the definition and the notation see Ref.~\cite[][Sec.~13.4]{Torquato2002}.}.
The reactants diffuse in the void phase and the spheres form perfectly absorbing traps~\footnote{For  packings of hard spheres, diffusion inside the spheres is obviously independent of the sphere configuration because the diffusion processes within the different spheres are isolated from one another.}.
In this diffusion-controlled regime, the time until the species diffuse to the trap interface is long relative to the characteristic time associated with the surface reaction~\footnote{The local mass conservation equation is a Laplace equation with Dirichlet boundary conditions}.
Figure~\ref{fig:visualization-diffusion} visualizes such a diffusion between perfectly absorbing spheres.
In steady state, the rate of removal by the traps is exactly compensated by production rate $G$ of reactants per unit volume (within the void phase).
The trapping constant $\gamma$ relates the average concentration field $C$ to the rate of production $G$~\footnote{In homogenization theory, the trapping constant is the average scaled auxiliary concentration field that solves the Laplace equation with Dirichlet boundary conditions~\cite{Torquato2002}.}
\begin{align}
  G = \gamma\mathcal{D}C,
\end{align}
where $\mathcal{D}$ is the diffusion constant.
The trapping constant is trivially related to the mean survival time $\tau$ of a Brownian particle.
It is the expected time that the diffusing particle survives before it is absorbed by a trap~\footnote{It is also the inverse of the average trapping rate.}.
For diffusion in the void phase, it is given by~\cite{Torquato2002}:
\begin{align}
  \tau = \frac{1}{ (1-\phi) \mathcal{D}\gamma}.
  \label{eq:tau-def}
\end{align}

The permeability and the trapping constant (or equivalently the mean survival time) characterize quite different physical processes.
Both are closely related to (different) geometrical properties of the pores.
The permeability may be regarded, roughly speaking, as an effective pore channel area of the dynamically connected part of the pore space,
and the trapping constant is related to the average pore size~\cite{Torquato2002}.

If the transport occurs in parallel channels of constant cross section, the permeability and mean survival time are strictly related to each other~\footnote{In contrast to this situation, for a disconnected pore space, the permeability vanishes but not the mean survival time}.
For more general porous media, there is a cross-property relation, an upper bound on the permeability in terms of the mean survival time~\cite{Torquato1990}.
Taking advantage of the similar underlying mathematical structure, similar variational bounds can be derived for both quantities
that are based on the same geometrical properties, namely, integrals over void and interfacial correlation functions.

Here, we estimate these structure integrals for MRJ sphere packings and determine an upper bound on their permeability as well as on their mean survival time.
Moreover, we use universal scalings to derive precise estimates for these quantities.
We compare these results to those for equilibrium hard-sphere liquids and overlapping spheres, as well as lattices formed by hard spheres.
Using the analytical bounds, we provide a prediction of the qualitative trend between these different structures without the need for time-consuming direct computer simulations.

\subsection{Bounds on the fluid permeability}

In the second paper of this series, we determined the two-point correlation function $S_2(r)$ of MRJ sphere packings, which is the probability that two random points at a distance $r$ lie in the phase formed by the spheres.
It is closely related to the void-void correlation function $F_{vv}$ and the autocovariance $\chi_{_V}(\vec{r})$ of the two-phase medium:
\begin{align}
  \chi_{_V}(\vec{r}) := F_{vv}(r) - (1-\phi)^2 = S_2(\vec{r}) - \phi^2,
\end{align}
where $\phi$ is the packing fraction.

Variational principles allows one to derive rigorous upper bounds on the fluid permeability~\cite{Torquato2002}.
Using an energy representation that is minimized by the (true) fluid permeability,
upper bounds can be derived from appropriate trial velocity fields (that satisfy the Stokes momentum equation without satisfying the incompressibility and no-slip conditions).
The following ``void" upper bound on $k$ in three dimensions is given by an integral over the autocovariance~\cite{Torquato2002}:
\begin{align}
  k/k_s \le \frac{12}{\phi D^2} \int_{0}^{\infty}r [S_2(r)-\phi^2]\mathrm{d}r.
  \label{eq:void-bound-permeability}
\end{align}
The normalizing constant $k_s=D^2/(18\phi)$ is  the Stokes dilute-limit permeability for a sphere of diameter $D$.

For overlapping spheres, the improper integral in Eq.~\eqref{eq:void-bound-permeability} and thus the bound on the fluid permeability can be calculated
using a numerical integration of the explicit analytic formulas for the two-point correlation function.
For the hard-sphere packings, a careful estimation of the improper integrals based on finite samples is needed.
We describe the details of our analysis in Appendix~\ref{sec_integration_S2}.

We find that the fluid permeabililty of MRJ sphere packings is bounded from above by $0.05965(8)k_s$,
which is less than half of the upper bound for the equilibrium hard-sphere liquid is given by $0.13090(6)k_s$.
In part this can be explained by the larger packing fraction of the MRJ spheres ($\phi=0.636$), but not entirely since MRJ spheres, unlike equilibrium hard spheres, are hyeruniform.
Comparing the bounds for overlapping spheres at $\phi=0.636$ to those at $\phi = 0.478$,
there is only a decrease by about $0.622$.

\begin{figure}[t]
  \centering
  \includegraphics[width=\linewidth]{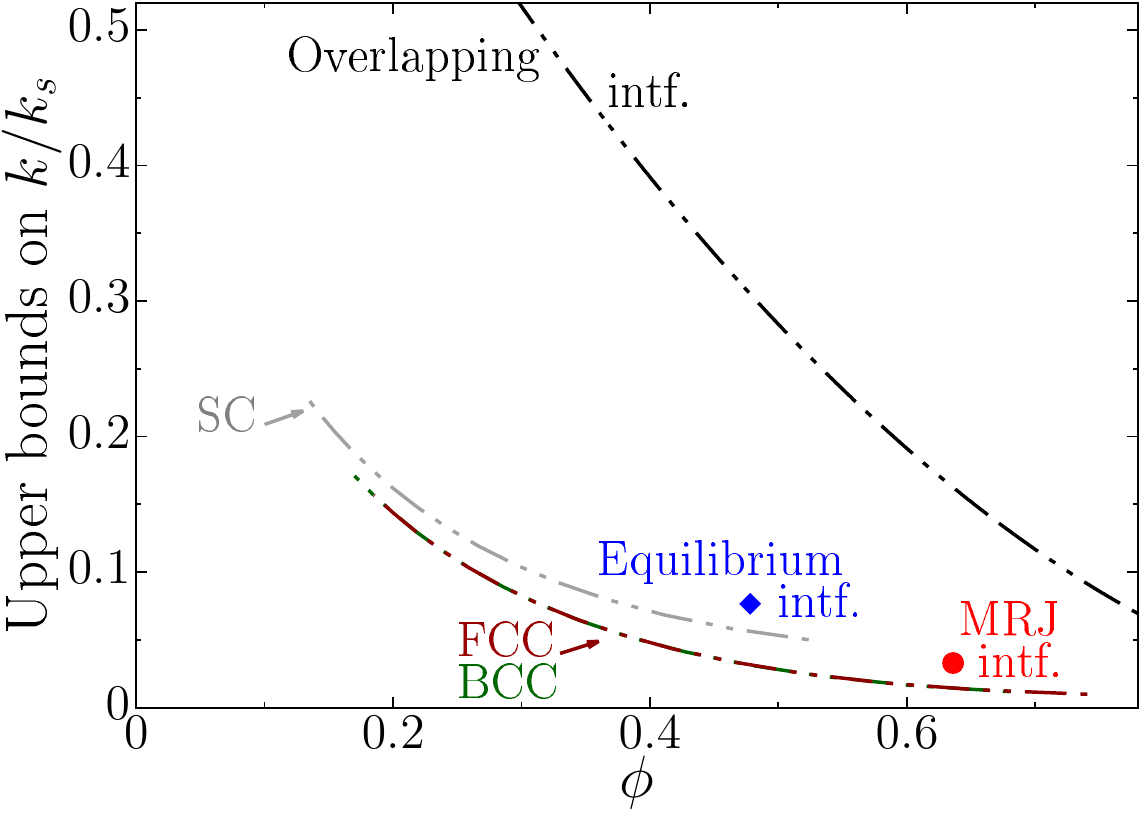}
  \caption{(Color online) The interfacial bounds on the fluid permeability $k$ as a function of the packing fraction $\phi$:
    The bounds for the MRJ sphere packings (at $\phi = 0.636$) are compared to those of an equilibrium hard-sphere liquid (at $\phi=0.478$) and overlapping spheres, as well as cubic lattices of hard spheres: the simple cubic (SC, gray lines), the body-centered cubic (BCC, green lines), and the face-centered cubic (FCC, dark-red lines) lattice.
    Because of the rescaling with $k_s=D^2/(18\phi)$, the bounds converge for $\phi\rightarrow 0$ to finite values.}
  \label{fig_void_bounds_permeability}
\end{figure}

Using variational principles, \citet{Doi1976} derived an upper bound on the fluid permeability that includes interfacial two-point information.
These bounds were rederived by Rubinstein and Torquato~\cite{RubinsteinTorquato1989JFM} and generalized to arbitrary dimension in Ref.~\cite{Torquato2002}.
The ``interfacial upper bound'' on the fluid permeability in three dimensions is given by
\begin{align}
  \begin{aligned}
  \frac{k}{k_s} \leq \frac{12\phi}{D^2}  \int_{0}^{\infty}r \Big[&\frac{(1-\phi)^2}{s^2}F_{ss}(r)\Big.\\
  & \Big.-\frac{2(1-\phi)}{s}F_{sv}(r)+F_{vv}(r)\Big]\mathrm{d}r,
  \end{aligned}
  \label{eq:interfacial-bound-permeability}
\end{align}
where $F_{ss}$ is the surface-surface correlation function, and $F_{sv}$ the surface-void correlation function; for more details, see the second paper of this series~\cite{KlattTorquato2016}.

For overlapping spheres, the improper integral can again be computed by a numerical integration of the explicit well-known formulas for the surface correlation functions.
For the estimation of the bounds for hard-sphere packings, see Appendix~\ref{sec_integration_S2}.
The resulting curves are shown in Fig.~\ref{fig_void_bounds_permeability}.

The additional information about the correlation of the interface strongly improves the bounds.
The upper bound on the fluid permeability decreases for the MRJ spheres by a factor of $0.551$ to $0.03287(7)k_s$,
and for the equilibrium hard-sphere liquid by a factor $0.585$ to $0.07661(8)k_s$.

Our estimate for the equilibrium hard-sphere liquid, which is based on an analytic integral of the explicit expressions for the correlation functions of finite simulated samples,
agrees with previous results~\cite{Torquato1986bounds} based on a Verlet-Weis modification~\cite{VerletWeis1972} of the Percus-Yevick approximation~\cite{HansenMcDonald2006}.

The complex structure of the MRJ packings, which suppresses large pores and pore channels, can be expected to result in a large trapping constant but small permeability,
as our calculation of the bounds suggests.

\subsection{Universal scaling of the fluid permeability}

\citet{MartysEtAl1994} found a universal scaling of the fluid permeability for sphere packings.
Using numerical simulations of Stokes flow and a scaling ansatz motivated by the rigorous bounds, they found excellent agreement for a variety of model microstructures.

Here we employ the following approximation of the permeability [based on Eq.~(6) in Ref.~\cite{MartysEtAl1994}]:
\begin{align}
  k \approx \frac{2(\phi + \phi_1^c)}{s^2}(1-\phi_2-\phi_1^c)^f,
\end{align}
where $s$ is the specific surface area, $f=4.2$, and $\phi_1^c = 0.0301(3)$ is the void percolation threshold of overlapping monodisperse spheres from Ref.~\cite{Rintoul2000}.

\begin{figure}[t]
  \centering
  \includegraphics[width=\linewidth]{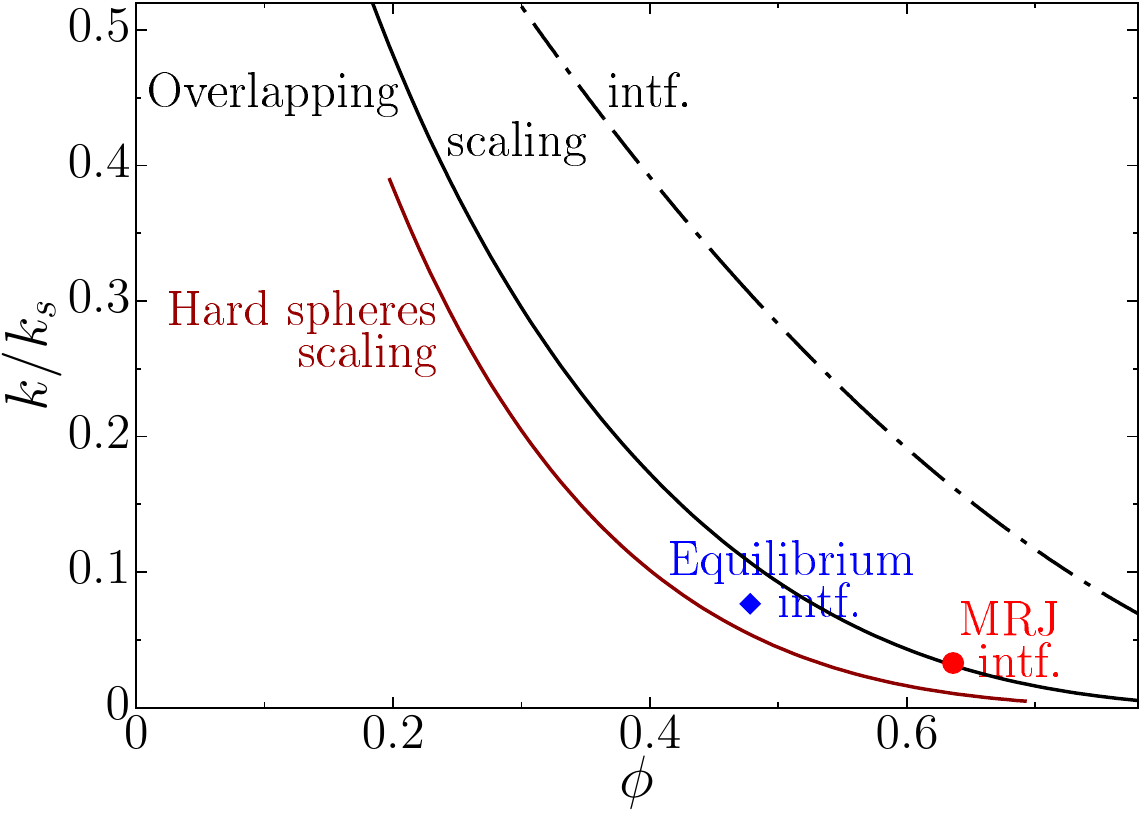}
  \caption{(Color online) Universal scaling of the fluid permeability $k$:
  The interfacial bounds for overlapping and equilibrium spheres as well as MRJ sphere packings are compared to the universal scaling law for hard and overlapping spheres (solid lines, dark-red and black).}
  \label{fig:permeability}
\end{figure}

In Fig.~\ref{fig:permeability}, the interfacial bounds for the disordered sphere packings (MRJ, equilibrium, and overlapping) are compared to the prediction of the universal scaling.
As expected, the bounds are not sharp, but they provide accurate trends when comparing different systems.

\subsection{Bounds on the mean survival time}

The integral from Eq.~\eqref{eq:void-bound-permeability} also provides a rigorous upper bound on the mean survival time
in three dimensions, which can again be derived by a variational principle~\cite{TorquatoRubinstein1989}:
\begin{align}
  {\tau}/{\tau_s} \leq \frac{12}{(1-\phi)\phi D^2}   \int_{0}^{\infty}r [S_2(r)-\phi^2]\mathrm{d}r.
  \label{eq:void-bound-trapping}
\end{align}
The normalizing constant $\tau_s:=D^2/(12\phi\mathcal{D})$ is the Smoluchowski result for the dilute-limit of a spherical trap of diameter $D$.
Note that the upper bounds on $\tau/{\tau_s}$ and $k/k_s$ only differ in a prefactor involving ($1-\phi$).

As mentioned in the preceding section, the same variational principles that provide a sharper interfacial bound on the 
fluid permeability also allow for an interfacial upper bound on the mean survival time~\cite{Doi1976, RubinsteinTorquato1988}.
Both of these bounds are based on the same improper integral of the surface two-point correlation functions.
They again only differ in the prefactor:
\begin{align}
  \begin{aligned}
    \frac{\tau}{\tau_s} \leq \frac{12\phi}{(1-\phi)D^2} \int_{0}^{\infty}r \Big[&\frac{(1-\phi)^2}{s^2}F_{ss}(r)\Big.\\
  & \Big.-\frac{2(1-\phi)}{s}F_{sv}(r)+F_{vv}(r)\Big]\mathrm{d}r.
  \end{aligned}
  \label{eq:tau-upper-bound}
\end{align}

The upper bound on $\tau$ can be complemented by a lower bound that is based on the mean size of pores in the void phase.
A pore size $\delta$ is the maximum radius of a spherical pore that can be assigned to a random point in the void phase such that the pore remains in the void phase; for more details, see the second paper of this series.
Given the mean pore size $\langle\delta\rangle$, that is, the first moment of the pore-size distribution, the mean survival time in the diffusion-controlled limit ($\kappa\rightarrow \infty$) in three dimensions is bounded from below by
\begin{align}
  \tau/\tau_s \geq \frac{12\phi\langle\delta\rangle^2}{D^2}.
  \label{eq:tau-lower-bound}
\end{align}
The bound was derived by \citet{Prager1963}.
A generalization for arbitrary absorbing traps, that is, finite values of $\kappa$, can be found in Ref.~\cite{TorquatoAvellaneda1991}.

The mean pore size for MRJ sphere packings was determined in the second paper of this series.
The resulting lower bound is $0.003990(2)D^2/\mathcal{D}$.
The bound for the MRJ spheres is by an order of magnitude smaller than the bound for overlapping spheres at the same volume fraction covered by spheres.
The latter bound is $0.0132977\cdots D^2/\mathcal{D}$.

For comparison, the lower bound for the equilibrium hard-sphere liquid at $\phi=0.478$ is $0.00968(1)D^2/\mathcal{D}$.
For equilibrium hard-sphere liquids at packing fractions below the freezing point, the pore-size bound can be calculated using analytic approximations of the pore size probability density~\cite[][Sec. 5.2.5.]{Torquato2002}.
Our simulation result is in good agreement with the analytic approximation.

\begin{figure}[t]
  \centering
  \includegraphics[width=\linewidth]{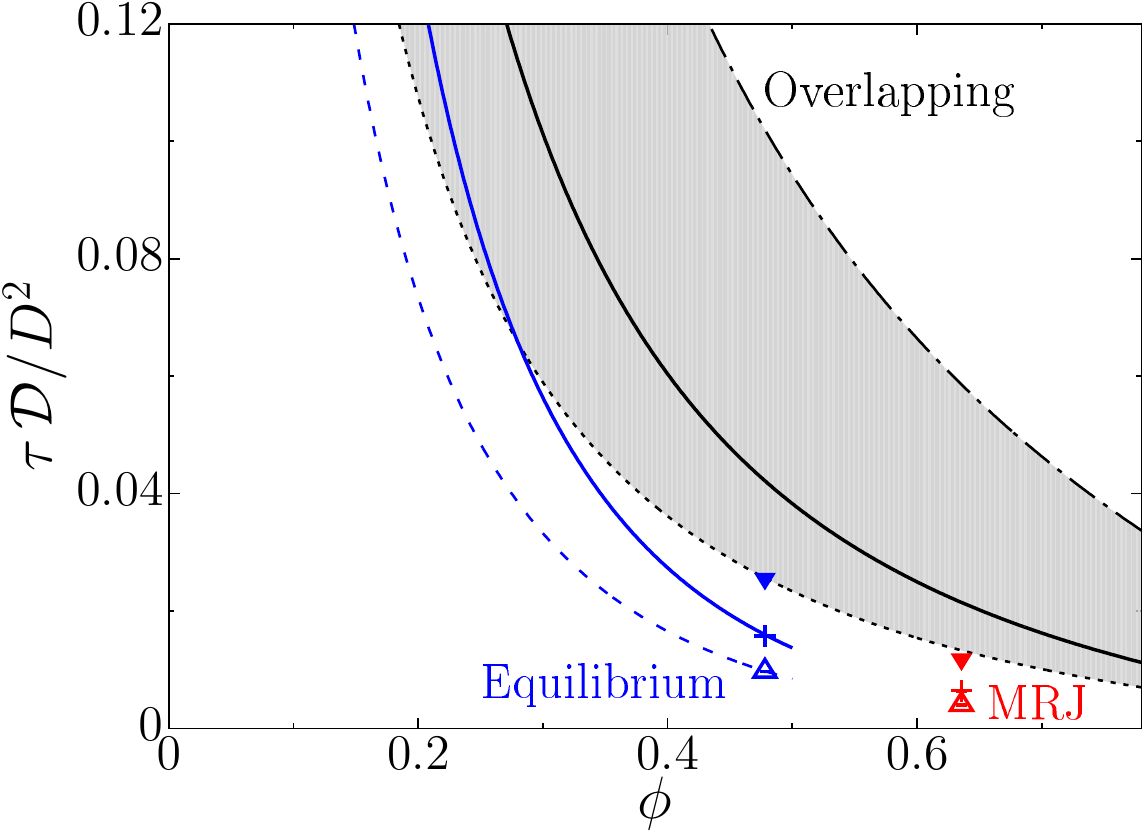}
  \caption{(Color online) Mean survival time $\tau$ rescaled by the diffusion constant $\mathcal{D}$ and the diameter $D$ of a single sphere and plotted as a function of the packing fraction $\phi$,
  is compared for the MRJ sphere packings (red), to equilibrium hard-sphere liquids (blue), and overlapping spheres (black).
  The interfacial upper bound, see Eq.~\eqref{eq:tau-upper-bound}, is indicated by closed triangles $\left(\blacktriangle\right)$ and a dashed-dotted line;
  the pore-size lower bound, see Eq.~\eqref{eq:tau-lower-bound}, by open triangles $\left(\triangle\right)$ and dashed lines.
  The gray-shaded region indicates for the overlapping spheres the mean survival times between these bounds.
  The solid lines and the crosses $\left(+\right)$ indicate the prediction of the mean survival time using the universal scaling relation, see Eq.~\eqref{eq:universal-scaling}.}
  \label{fig:mean-survival-time}
\end{figure}

Figure~\ref{fig:mean-survival-time} depicts both the upper and lower bounds on the mean survival time; cf. Fig.~\ref{fig_void_bounds_permeability} for the fluid permeability.


\subsection{Universal scaling of the mean survival time}
\label{sec:scaling-law}

By rescaling the mean survival time by the diffusion constant $\mathcal{D}$ and the diameter $D$ of a single sphere,
Torquato and Yeong~\cite{TorquatoYeong1997} found a universal scaling law for a broad class of model particulate- and digitized-based models.
It allows for accurate predictions of $\tau$ based on the mean pore size, specific surface, and porosity.

First, a characteristic time scale $\tau_0$ is defined:
\begin{align}
  \tau_0 := \frac{3\phi}{s^2(1-\phi)\mathcal{D}},
\end{align}
where $s$ is the specific surface of the two-phase medium, that is, the ratio of the surface area and the volume of the whole system; for more details, see the second paper of this series.

The universal scaling of the mean survival time is then given by
\begin{align}
  \frac{\tau}{\tau_0} = \frac{8}{5} \frac{\langle\delta\rangle^2}{\tau_0\mathcal{D}} + \frac{8}{7} \left(\frac{\langle\delta\rangle^2}{\tau_0\mathcal{D}}\right)^2.
  \label{eq:universal-scaling}
\end{align}
Thus the mean survival time can be predicted from the packing fraction $\phi$, specific surface $s$, mean pore size $\langle\delta\rangle$, and diffusion constant $\mathcal{D}$~\footnote{Using
the estimates of the mean pore size for each packing, which were already determined in the second paper of this series,
we predict the mean survival time with the universal scaling using the exact packing fraction, and determine the weighted mean.}.

Figure~\ref{fig:mean-survival-time} shows the predictions of the scaling law for the mean survival time $\tau$ for diffusion in the pores of MRJ sphere packings.
The results are compared to those of overlapping spheres and equilibrium hard-sphere liquids.
For the latter, the above mentioned approximation of the pore size probability density~\cite[][Sec. 5.2.5.]{Torquato2002} allows for an accurate prediction of the mean survival time of equilibrium hard-sphere liquids for packing fractions below the freezing transition.

For the MRJ sphere packings, the approximation of the mean survival time yields $0.006431(3) D^2/\mathcal{D}$.
It is indeed less than half of the mean survival time $0.01580(2) D^2/\mathcal{D}$ that is predicted for the equilibrium hard-sphere liquid (at $\phi=0.478$).
Moreover, it is more than three times smaller than the mean survival time $0.02146\ldots D^2/\mathcal{D}$ for overlapping spheres at the same packing fraction as the MRJ spheres (at $\phi=0.636$).

As discussed above and expected from the bounds, the strong suppression of large pores in the MRJ sphere packings dramatically reduces the mean survival time
compared to the equilibrium hard-sphere liquid and especially to the very irregular pattern of non-interacting spheres.


\section{Principal diffusion relaxation time}
\label{sec:mean-survival-time}

Both the trapping constant and the mean survival time characterize the steady-state trapping problem, that is, if the concentration does not change over time.
The time-dependent trapping problem with a continuous decay of the concentration is characterized by the associated relaxation times.
These time scales are closely related to the characteristic length scales of the pore region.
Besides chemical reactions, this process is also especially important for the nuclear magnetization density in nuclear magnetic resonance experiments.

In the diffusion controlled limit, the concentration obeys a Laplace equation, the time-dependent diffusion equation, with Dirichlet boundary conditions, that is, the concentration vanishes at the interface of the void and particle phase.
The corresponding eigenvalues $\lambda_n$ (with $0<\lambda_1\leq \lambda_2 \leq \ldots$) define the diffusion relaxation times
\begin{align}
  T_n := \frac{1}{\mathcal{D}\lambda_n}.
  \label{eq:def-Tn}
\end{align}
The principle diffusion relaxation time $T_1$ dominates the behavior at long times, for example, the tail of the survival probability is proportional to $\exp(-t/T_1)$.

\begin{figure}[t]
  \centering
  \includegraphics[width=\linewidth]{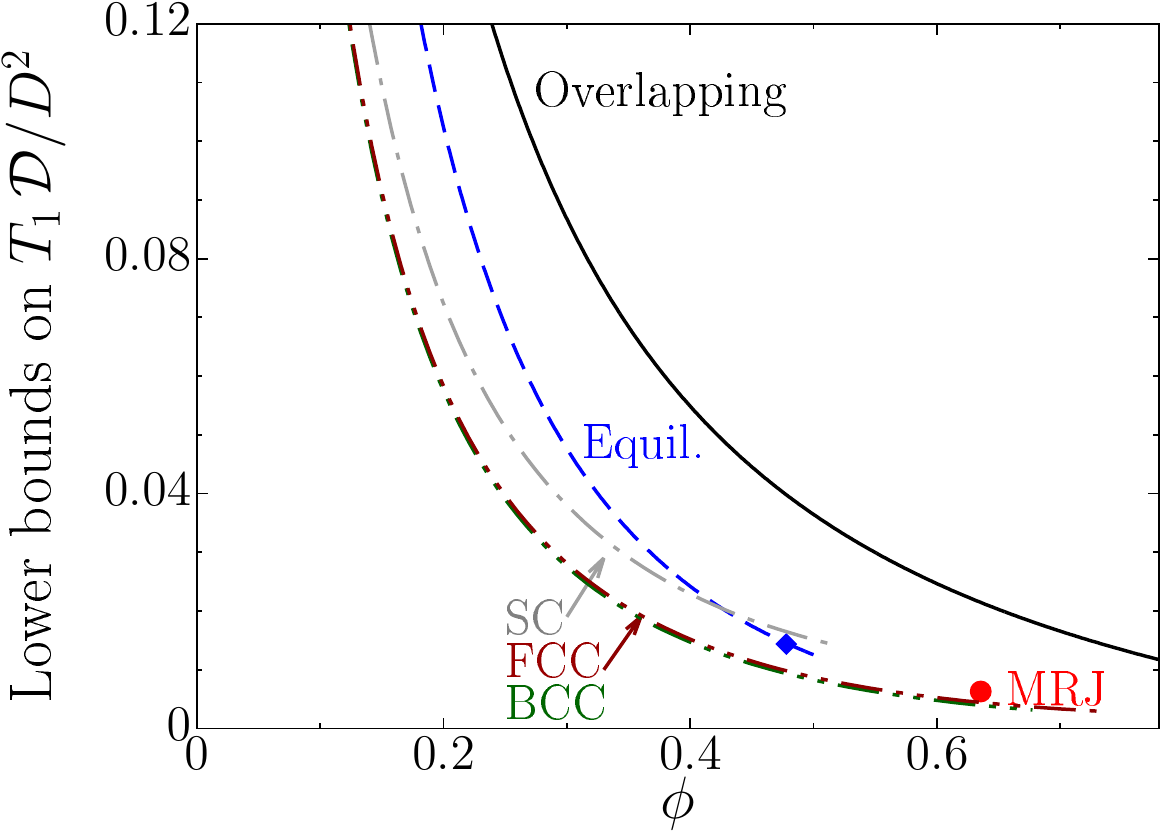}
  \caption{(Color online) Pore-size lower bounds on the principal diffusion relaxation time $T_1$, see Eqs.~\eqref{eq:def-Tn} and \eqref{eq:bound-T1}:
  the bounds for the MRJ sphere packings is distinctly smaller than those for equilibrium or overlapping spheres, and only slightly larger than dense lattice packings of hard spheres, which are perfectly ordered but anisotropic.}
  \label{fig:principal-relaxation-time}
\end{figure}

Like the mean survival time $\tau$, the principal relaxation time $T_1$ is intimately related to the first and second moments of the pore-size probability density function~\cite{TorquatoAvellaneda1991}.
While the mean pore sizes allows one to predict the mean survival time of the steady-state problem, the variance of the pore sizes provides a bound on the principal relaxation time in the time-dependent problem.
Based on the first and second moment of the pore size distribution, \citet{TorquatoAvellaneda1991} found a lower bound on the principal diffusion relaxation time $T_1$ for arbitrary values of the surface reaction rate $\kappa$ (using a variational principle similar to that used for the lower bound on the mean survival time).
In the here studied diffusion-controlled regime, that is, in the limit $\kappa\rightarrow\infty$ where a particle is immediately trapped once it diffuses to the surface, the lower bound is given by only the second moment of the pore-size distribution:
\begin{align}
  T_1 \geq \frac{\langle \delta^2 \rangle}{\mathcal{D}}.
  \label{eq:bound-T1}
\end{align}

Like the first moment of the pore sizes in MRJ sphere packings, we have also determined the second moment in the second paper of this series and compared it to that of the equilibrium hard-sphere liquid.
The lower bound for the MRJ spheres is $0.006301(3)D^2/\mathcal{D}$, which is less than half of the corresponding value for the equilibrium hard-sphere liquid $0.01437(2)D^2/\mathcal{D}$.
The results are shown in Fig.~\ref{fig:principal-relaxation-time},
which includes also the lower pore-size bound for equilibrium hard-sphere liquids as a function of the packing fraction, based on the analytic approximation of the pore size probability density~\cite[][Sec. 5.2.5.]{Torquato2002}.
The principal relaxation time can be expected to be distinctly smaller in the hyperuniform MRJ state than in the more irregular equilibrium fluid, and close to the crystalline (but anisotropic) BCC and FCC structures.


\section{Effective dielectric constant}
\label{sec:eff_dielec}

The dielectric constant, or permittivity, relates the polarization in a medium to the applied electric field.
In heterogeneous materials, the propagation of electromagnetic waves can be described by an effective complex dielectric tensor~$\bm{\varepsilon}_e$.
It is needed, for example,
for remote sensing~\cite{tsang_theory_1985},
to study wave propagation through turbulent atmospheres~\cite{tatarskii1971effects},
to actively manipulate composites~\cite{meola_non-destructive_2002},
to probe artificial materials~\cite{PhysRevB.50.15636},
and especially to study electrostatic resonances~\cite{McPhedran1980, doi:10.1063/1.2718279}.

\begin{figure}[t]
  \centering
  \includegraphics[width=\linewidth]{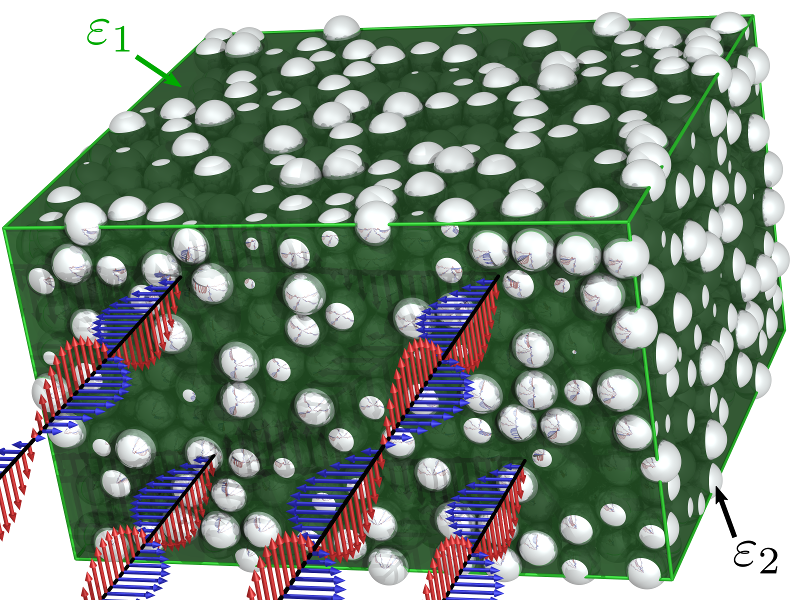}
  \caption{(Color online) Schematic of a plane electromagnetic wave in the long-wavelength limit (the red and blue arrows depict its values along four lines) entering a heterogeneous dielectric medium that is formed by an MRJ hard-sphere packing.
           The spheres (steel-gray) have a dielectric constant $\varepsilon_2$, and the intermediate space (green) has a dielectric constant $\varepsilon_1$.}
  \label{fig:visualization-em-waves}
\end{figure}

We are interested in the long-wavelength regime, that is, when the wavelength is much larger than the scale of the inhomogeneities of the two-phase medium (because in the complementary regime ray-tracing techniques yield appropriate solutions).
This is schematically depicted in Fig.~\ref{fig:visualization-em-waves}.
\citet{RechtsmanTorquato2008} derived an exact strong-contrast expansion for $\bm{\varepsilon}_e$, which is explicitly given in terms of integrals over the $n$-point correlation function of the random two-phase medium.
Because the heterogeneous microstructures studied here are isotropic, we can use a scalar effective dielectric constant~$\dc{e}$.
For former approximations of the effective dielectric constant see, \eg, Refs.~\cite{tsang_theory_1985, PhysRevB.50.15636, PhysRevB.47.8539}.

In a typical isotropic, locally disordered system there is dissipation for the transport of electromagnetic waves with long wavelengths (due to density fluctuations in the system).  
Dissipationless transport is only possible in crystal lattices with no large-scale density fluctuations.
However, crystals are not isotropic.
An amorphous hyperuniform system like the MRJ sphere packing can be both isotropic and globally uniform, similar to a crystal.
It could allow for an optimal solution among the isotropic systems.
Therefore, unique transport properties can be expected.

\begin{figure*}
  \centering
  \includegraphics[width=\linewidth]{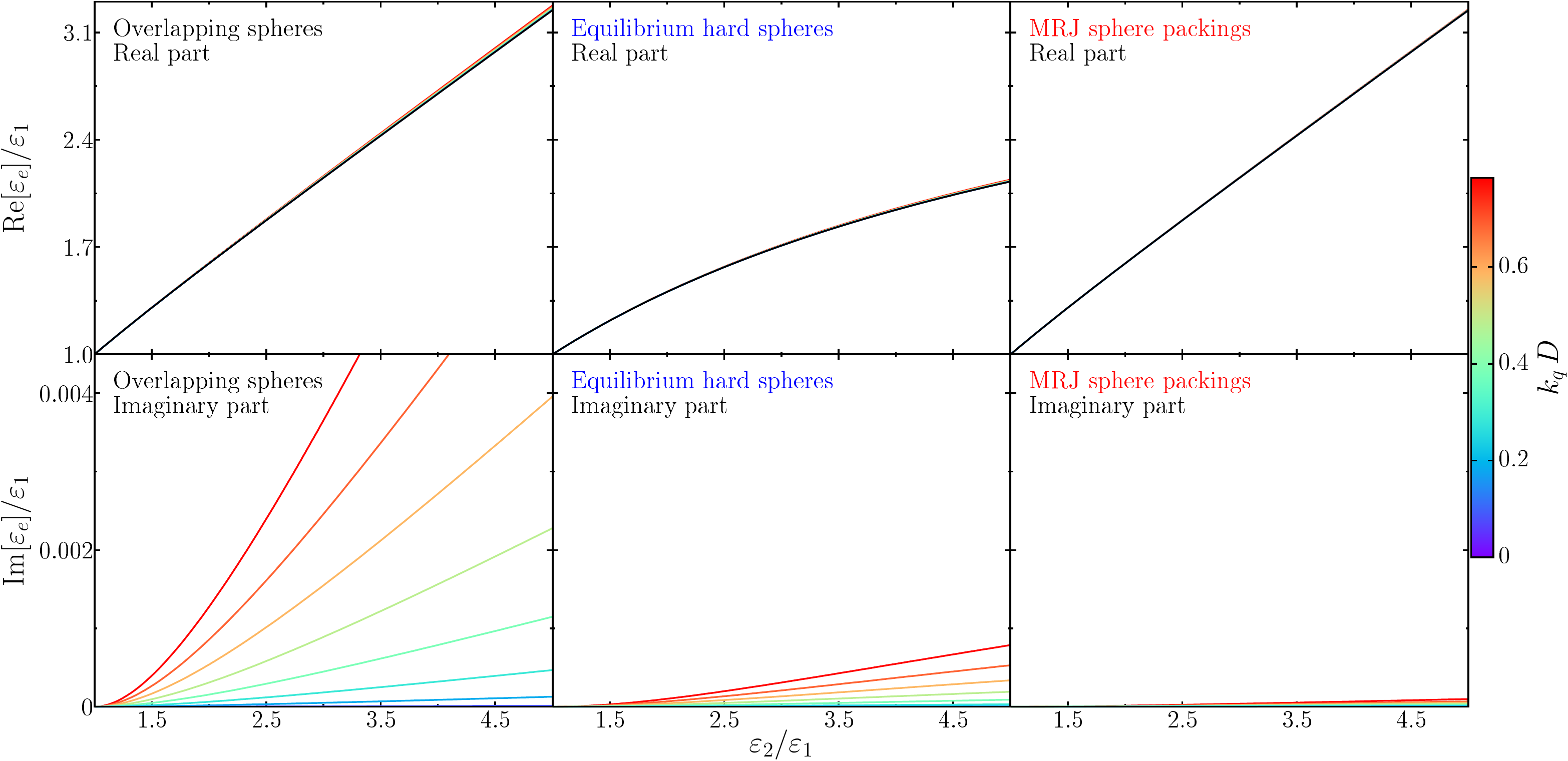}
  \caption{(Color online) The two-point approximation of the effective dielectric constant $\dc{e}$ in the strong-contrast expansion, see  Eq.~\eqref{eq:epsilon-e}, as a function of the ratio $\dc{2}/\dc{1}$ of the dielectric constants in the particle and void phase for various absolute values of the wave vectors $k_q$.
    Both the real part (top row) and imaginary part (bottom row) are plotted for the MRJ state (right column) and compared to the results for the equilibrium hard spheres (middle column) and
    overlapping spheres (left column).}
  \label{fig:dielectric}
\end{figure*}

\subsection{The two-point approximation of the strong-contrast expansion}
\label{sec:strong-contrast-expansion}

Following Ref.~\cite{Torquato2002}, \citet{RechtsmanTorquato2008} expanded the effective dielectric tensor~$\bm{\varepsilon}_e$ for a random two-phase medium where the single phases are assumed to be isotropic.
We here additionally assume that their dielectric constants are real valued.
Due to the disorder of the two-phase medium, the effective dielectric constant will nevertheless exhibit a non-vanishing imaginary part.

One of the two phases is chosen as a reference phase surrounding the heterogeneous material.
The phase of reference is here denoted by $q$, the complementary phase by $p$.
Both choices $q=1$ and $q=2$ yield valid strong-contrast expansions.
Depending on the microstructure, the rate of convergence may differ.

The strong contrast refers to the difference between the two dielectric constants $\dc{1}$ and $\dc{2}$ of the two phases $i=1,2$.
By choosing a beneficial strong-contrast expansion parameter $\beta_{pq}$, based on Pad\'e approximations, they achieved a wide radius of convergence (compared to the weak-contrast expansion that uses only the difference $\dc{1}-\dc{2}$).
Whenever one phase consists of disconnected (dispersed) particles in a percolating matrix, the expansion will converge rapidly~\cite{Torquato2002, RechtsmanTorquato2008} and hence truncation of the series at low orders can yield a very good approximation of the effective dielectric constant.
In three dimensions,
\begin{align}
  \beta_{pq} := \frac{\dc{p}-\dc{q}}{\dc{p}+2\dc{q}}.
\end{align}
Accordingly, the effective polarizability $\beta_{eq}$ is defined using the effective dielectric constant $\dc{e}$:
\begin{align}
  \beta_{eq} := \frac{\dc{e}-\dc{q}}{\dc{e}+2\dc{q}}.
\end{align}
This strong contrast expansion allows predictions for large ratios between the dielectric constants of the two phases.
The results of Ref.~\cite{RechtsmanTorquato2008} suggest that (in the long-wavelength limit) truncation of the expansion at the level of the two-point correlation function $S_2(r)$ still yields a good approximation to the effective dielectric constant for any phase contrast ratio for the aforementioned class of particle dispersions.
It is given by
\begin{align}
  \beta_{pq}\phi_p^2\beta_{eq}^{-1} & = \phi_p - \beta_{pq} A_2,
  \label{eq:two-point-dielectric}
\end{align}
where $A_2$ includes the two-point information about the random media:
\begin{align}
  A_2 := 2k_q^2 \int_0^{\infty} \text{e}^{ik_qr}r[S_2^{(p)}(r)-\phi_p^2]dr,
  \label{eq:def-A2}
\end{align}
using the two-point correlation function $S_2^{(p)}(r)$ and volume fraction $\phi_p$ of phase $p$.

In the static problem, $A_2$ always vanishes for a statistically isotropic and homogeneous medium.
For the propagation of electromagnetic waves, the imaginary part of $A_2$ corresponds to its attenuation in the disordered medium, see Ref.~\cite{RechtsmanTorquato2008}.

In the long-wavelength limit, it is determined in leading-order by the local-volume-fraction fluctuations.
More precisely, it is proportional to the square of the coarseness of the material, that is, of the asymptotic standard deviation of volume fractions in the limit of an infinitely large window.
As a rule of thumb, the more disordered a heterogeneous material, the stronger the attenuation of the electromagnetic waves.
Interestingly, the coarseness vanishes, by definition, for a hyperuniform material.
A hyperuniform material allows for an almost dissipationless transport of electromagnetic waves.

Formally, this can be shown by the Taylor expansion of $A_2$ through third order in $k_q$ at $k_q=0$:
\begin{align}
  \begin{aligned}
  A_2 & =   2k_q^2 \int_0^{\infty} r[S_2^{(p)}(r)-\phi_p^2]dr \\
     & \qquad + 2ik_q^3 \int_0^{\infty} r^2[S_2^{(p)}(r)-\phi_p^2]dr + \mathcal{O}(k_q^4).
  \end{aligned}
\end{align}
The leading term is proportional to the spectral density at wavenumber zero $\tilde{\chi}_{_V}(0)$, see Eq.~(11) in the second paper of this series~\cite{KlattTorquato2016}. 
A two-phase medium is hyperuniform if and only if $\tilde{\chi}_{_V}(0)=0$.
Therefore, the leading term of $A_2$ vanishes for a hyperuniform material.
The material is is exactly dissipationless through lowest order in perturbation expansion and third order in $k_q$.
We can therefore expect a small attenuation for the MRJ sphere packing.

Note that $A_2$ at finite values of $k_q$ is proportional to the spectral density $\tilde{\chi}_{_V}(k)$.
For a stealthy disordered hyperuniform system, the spectral density vanishes for all wave vectors below a finite threshold $k<k_0$~\cite{TorquatoZhangStillinger2015}.
In such a system, a completely dissipationless transport of electromagnetic waves with long but finite wavelengths is possible.

\subsection{The effective dielectric constant of MRJ sphere packings}

Using the second-order approximation from Eq.~\eqref{eq:two-point-dielectric}, we here estimate both the real and imaginary parts of the effective dielectric constant $\dc{e}$ for the MRJ sphere packings ($\phi=0.636$).
We compare them to those of the hard-sphere liquid ($\phi=0.478$) and overlapping spheres (at the same mean volume fraction as the MRJ packings).
For the MRJ sphere packings as well as for the overlapping spheres, the particle phase is chosen as the phase of reference, \ie, $q=2$.
For the equilibrium hard-sphere liquid, the particle phase does not percolate.
Therefore, the void phase serves as the phase of reference, \ie, $q=1$.
These choices for the overlapping and equilibrium spheres agree with those in Ref.~\cite{RechtsmanTorquato2008}.

Solving Eq.~\eqref{eq:two-point-dielectric} for $\dc{e}$, we derive
\begin{align}
  \frac{\dc{e}}{\dc{q}} &= \frac{\phi_p-A_2\beta_{pq}+2\beta_{pq}\phi_p^2}{\phi_p-A_2\beta_{pq}-\beta_{pq}\phi_p^2},
  \label{eq:epsilon-e}
\end{align}
which provides a good approximation to all orders.
To calculate $A_2$, we numerically integrate the analytic two-point correlation function $S_2(r)$ from the second paper of this series.
We compute $A_2$ for each sample separately and then estimate the expectations of the real and imaginary parts of $A_2$.

Figure~\ref{fig:dielectric} shows the two-point approximation of both the real and imaginary part of $\dc{e}$ for the MRJ spheres, and compares them to those of the overlapping and equilibrium hard spheres.
The effective dielectric constant $\dc{e}$ is plotted as a function of the dielectric constant $\dc{2}$ of the particle phase, both in units of the dielectric constant $\dc{1}$ of the void phase.
In each plot, the family of curves represents various absolute values $k_q$ of the wave vectors, where $k_q=0$ represents the static case and $k_q \approx 0.7/D$ corresponds to a wavelength about nine times the diameter $D$ of a single sphere.

If $\dc{2} \rightarrow \dc{1}$, the two phases become indistinguishable and $\dc{e}\rightarrow \dc{1}$.
For $\dc{2} > \dc{1}$, the effective dielectric constant is strongly monotonically increasing as a function of $\dc{2}$.

The real part of $\dc{e}$ hardly depends on $k_q$, in contrast to the imaginary part.
For the overlapping spheres, the latter strongly increases even within the range of $k_q$ studied here.
As mentioned above, the integral $A_2$ vanishes in the static case ($k_q=0$) for any isotropic system and therefore the imaginary part of $\dc{e}$ vanishes.

Moreover, in the static case, the real part of $\dc{e}$ only depends only on the volume fraction $\phi$ (because of $A_2=0$).
Therefore, $\text{Re}[\dc{e}]$ is identical for the overlapping spheres and the MRJ state despite their extremely different microstructure.
Although it depends only weakly on $k_q$ and the difference between $\text{Re}[\dc{e}]$ for overlapping or for MRJ spheres remain small even for $k_q\approx0.7/D$,
we observe that the real part of the dielectric constant remains smaller for the MRJ than for the overlapping spheres.
An explanation is that the MRJ spheres are less connected.

In the dynamic case ($k_q>0$), the imaginary part $\text{Im}[\dc{e}]$ depends distinctively on subtle structural details.
In accord with Ref.~\cite{RechtsmanTorquato2008}, the imaginary part is much larger for the overlapping spheres than for the equilibrium hard spheres.
This is intuitive, because the non-overlap constraint in the latter system decreases the coarseness of the system (compared to the completely independent overlapping spheres).

Most interestingly, we find an even much smaller imaginary part of the dielectric constant for the MRJ state.
It practically vanishes for a considerable range of phase contrast ratio and wavelengths.
This vanishing of $\text{Im}[\dc{e}]$ arises from the hyperuniformity of the MRJ sphere packings, as predicted in Sec.~\ref{sec:strong-contrast-expansion}.
The coarseness converges to zero for $k_q\rightarrow 0$;
in the second paper of this series, we showed how the spectral density $\tilde{\chi}_{_V}$ vanishes in the limit of infinite wavelengths.
The small functional values of $\tilde{\chi}_{_V}$ at small absolute values of the wave vector are, as mentioned above, strictly related to small leading-order terms of $\text{Im}[\dc{e}]$, see Ref.~\cite[][Eq. (B5)]{RechtsmanTorquato2008}.
Roughly speaking, propagating waves in the long-wavelength limit are no longer scattered off because the heterogeneities vanish on large length scales.
Maximally random jammed spheres form a nearly dissipationless disordered and isotropic two-phase medium.


\section{Micro- and macroscopically anisotropic packings of spheroids}
\label{sec_spheroids}

\begin{figure}[t]
  \centering
  \includegraphics[width=\linewidth]{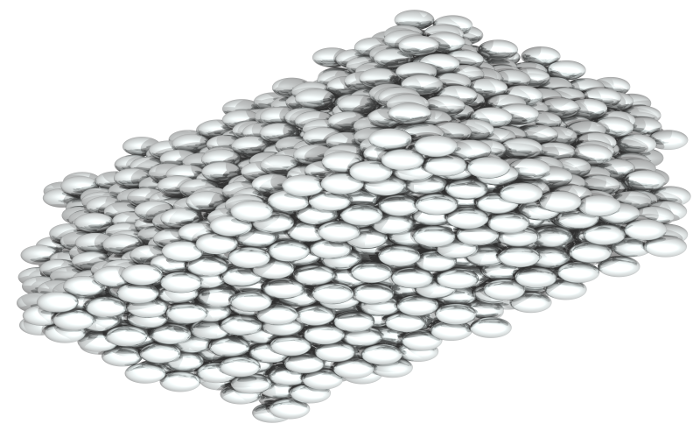}
\caption{An anisotropic packing of spheroids resulting from a linear transformation (stretching) of an MRJ sphere packing.}
  \label{fig:packing-spheroids}
\end{figure}

Hitherto, we have examined two-phase media composed of an isotropic distribution of spheres in a matrix phase.
By a linear transformation, a scaling in one direction, they can easily be generalized to anisotropic packings of oriented spheroids, see Fig.~\ref{fig:packing-spheroids}.
There is a scaling transformation that maps the results of the isotropic into equivalent results for the anisotropic systems~\cite{LebowitzPerram}.
Both microstructural and physical properties of the oriented spheroids can be expressed by properties of the corresponding sphere systems~\cite{sen_effective_1989, torquato_conductivity_1990, lado_twopoint_1990, torquato_trapping_1991}.

If slight deviations from the perfect alignment appear in experimental realizations, 
the rigorous bounds for perfectly aligned grains should still be a good approximation.
Of course, in the opposite case of an isotropic distribution of orientations, the effective properties could differ dramatically, especially for highly aspherical shapes.
Then the system would be macroscopically isotropic, i.e., the conductivity tensor is isotropic.

Here we use the explicit bounds on (anisotropic) physical properties of aligned spheroids from Ref.~\cite{torquato_trapping_1991}.
Such a system can model certain laminates or short-fiber composites.

The symmetry axis of the spheroids is chosen to point in $z$-direction, and the corresponding semi axis has length $b$.
The other semi-axis length $a$ is equal to the radius of the original spheres.
The aspect ratio is denoted by $\epsilon=b/a$; $\epsilon = 0$ corresponds to disks and $\epsilon=\infty$ to needles.

\subsection{Mean survival time}

\begin{figure}[t]
  \centering
  \includegraphics[width=\linewidth]{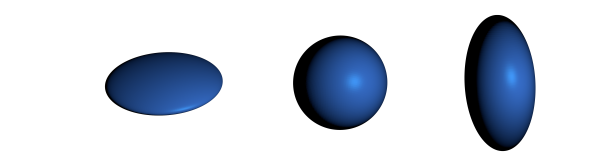}\\
  \includegraphics[width=\linewidth]{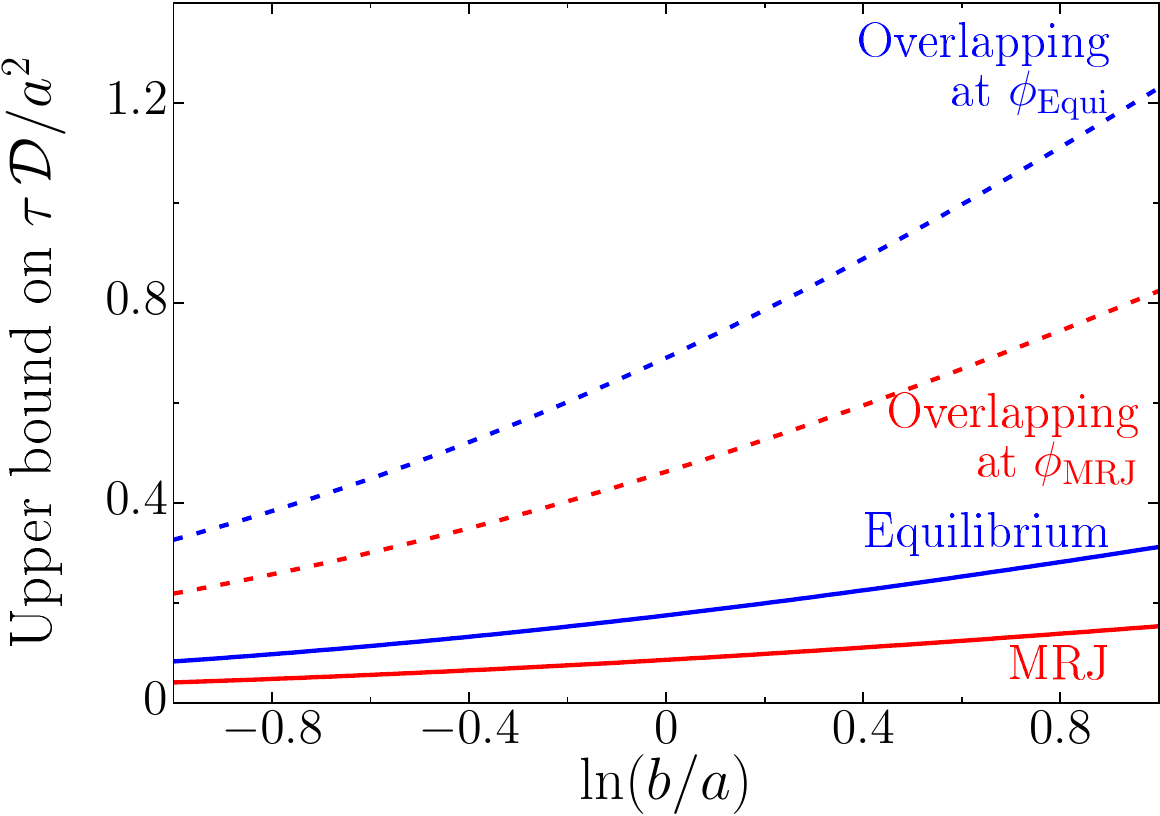}
\caption{(Color online) Upper bound on the mean survival time $\tau$ as a function of the aspect ratio $\epsilon = b/a$ of the spheroids. At the top, exemplary spheroids are depicted; from left to right: oblate ellipsoid ($\epsilon < 1$), sphere ($\epsilon=1$), and a prolate ellipsoid ($\epsilon >1$).}
  \label{fig:tau-spheroids}
\end{figure}

Using the scaling transformation, \citet{torquato_trapping_1991} generalized the two-point upper bound
on the mean survival time of a particle diffusing between spheres~\cite{TorquatoRubinstein1989}, see Eq.~\eqref{eq:void-bound-trapping},
to the corresponding bound for oriented spheroids.
The latter is simply given by the bound for spheres multiplied by a factor $\xi_1(\epsilon)$, which only depends on the aspect ratio $\epsilon$ of the spheroids:
\begin{align}
  \xi_1(\epsilon) = 
  \begin{cases}
    \frac{\epsilon}{\sqrt{1-\epsilon^2}} \tan^{-1}\frac{\sqrt{1-\epsilon^2}}{\epsilon}, & \epsilon < 1 \text{ (oblate)} \\
    \frac{\epsilon}{2\sqrt{\epsilon^2-1}}\ln \frac{\epsilon+\sqrt{\epsilon^2-1}}{\epsilon-\sqrt{\epsilon^2-1}}, & \epsilon > 1 \text{ (prolate)}.
  \end{cases}
\end{align}
The bounds for MRJ and equilibrium sphere packings (as well as for overlapping spheres at the same volume fractions) are plotted in Fig.~\ref{fig:tau-spheroids}.
As expected, the bound on $\tau$ increases monotonically with the aspect ratio, and the overlapping spheroids have a larger mean survival time than the hard-spheroid packings at the same volume fraction,
because a smaller surface area available for reaction increases the mean survival time~\cite{torquato_trapping_1991}.

\subsection{Effective conductivity}

Electric or thermal conductivity are mathematically equivalent.
In an isotropic material, the conductivity is a scalar proportionality constant between the electric current and the electric field (Ohm's law) or between the heat flux and the temperature gradient (Fourier's law).
If the material is anisotropic, the conductivity is a tensor, a linear mapping of the electric current to the field (or of the heat flux to the temperature gradient).
For a macroscopically anisotropic material, the conductivity tensor $\bm{\sigma}$ is not proportional to the unit tensor, but there will be a preferred orientation of the flux.

The MRJ sphere packings are an example of a ``bicontinuous'' two-phase system~\cite{Torquato2002}, that is, they have a both a percolating particle phase (due to the rigid, jammed backbone) and a percolating matrix phase.
Thus, if the phase conductivity contrast ratio of particles to matrix is infinite (e.g., for superconducting spheres), then the effective conductivity of the two-phase system will be superconducting.
We note in passing that the percolation behavior of MRJ sphere packings have recently been studied~\cite{ZiffTorquato2017}.

For macroscopically isotropic two-phase media, \citet{HashinShtrikman} derived the optimal bounds on the effective conductivity given only the volume fractions of the phases.
Actually, these are two-point bounds, but because they only depend on the two-point correlation function in the limits $r\rightarrow 0$ and $r\rightarrow \infty$, only the volume fractions are explicitly involved.

Their anisotropic generalization for the effective conductivity tensor $\bm{\sigma_e}$ depends nontrivially on the two-point correlation function if there are no further symmetry assumptions~\cite{sen_effective_1989, torquato_conductivity_1990, torquato_trapping_1991}.
In the special case of aligned grains with transverse isotropy and azimuthal symmetry (like our spheroids), the two-point bounds depend only on the volume fraction explictly, like in the isotropic case.

Without loss of generality, it is therefore sufficient to state the bounds for $\sigma_2 \geq \sigma_1$, where $\sigma_1$ and $\sigma_2$ are the bulk conductivities of phases 1 and 2, respectively.
Simply by renaming the phases, the bounds for $\sigma_1 \geq \sigma_2$ follow; this results in a trivial ``swapping'' of upper and lower bounds.

For our choice of spheroids aligned in $z$-direction, the effective conductivity tensors can be represented by a diagonal matrix:
\begin{align}
  \bm{\sigma_e} = \begin{bmatrix}
  (\bm{\sigma_e})_{xx} & 0 & 0\\
  0 & (\bm{\sigma_e})_{yy} & 0\\
  0 & 0 & (\bm{\sigma_e})_{zz}
  \end{bmatrix},
\end{align}
where $(\bm{\sigma_e})_{xx}=(\bm{\sigma_e})_{yy}$ but $(\bm{\sigma_e})_{xx}\neq (\bm{\sigma_e})_{zz}$ for $\epsilon \neq 1$. 

The upper bound $(\bm{\sigma_U})_{xx}$ and lower bound $(\bm{\sigma_L})_{xx}$ on $(\bm{\sigma_e})_{xx}$ are given by~\cite{sen_effective_1989, torquato_conductivity_1990, torquato_trapping_1991}:
\begin{align}
  \label{eq:boundsigma-A}
  \frac{(\bm{\sigma_U})_{xx}}{\sigma_1} &= \frac{\sigma_2}{\sigma_1} + \frac{\phi_1(1-\frac{\sigma_2}{\sigma_1})}{1+\phi_2 Q \frac{1-\sigma_2/\sigma_1}{\sigma_2/\sigma_1}},  \\
  \frac{(\bm{\sigma_L})_{xx}}{\sigma_1} &= 1 + \frac{\phi_2(\frac{\sigma_2}{\sigma_1}-1)}{1+\phi_1 Q (\frac{\sigma_2}{\sigma_1}-1)},
\end{align}
and the corresponding bounds $(\bm{\sigma_U})_{zz}$ and lower bound $(\bm{\sigma_L})_{zz}$ on $(\bm{\sigma_e})_{zz}$ are given by~\cite{sen_effective_1989, torquato_conductivity_1990, torquato_trapping_1991}:
\begin{align}
  \frac{(\bm{\sigma_U})_{zz}}{\sigma_1} &= \frac{\sigma_2}{\sigma_1} + \frac{\phi_1(1-\frac{\sigma_2}{\sigma_1})}{1+\phi_2 (1-2Q) \frac{1-\sigma_2/\sigma_1}{\sigma_2/\sigma_1}},  \\
  \frac{(\bm{\sigma_L})_{zz}}{\sigma_1} &= 1 + \frac{\phi_2(\frac{\sigma_2}{\sigma_1}-1)}{1+\phi_1 (1-2Q) (\frac{\sigma_2}{\sigma_1}-1)},
  \label{eq:boundsigma-B}
\end{align}
where $Q$ depends only on the shape of the single grains, \ie, on the aspect ratio $\epsilon$ of the spheroids:
\begin{align}
  Q = \frac{1}{2} + \frac{1}{2\epsilon^2 - 2} \left[ 1- \xi_1(\epsilon) \right].
\end{align}

\begin{figure}[t]
  \centering
  \begin{minipage}[t]{0.07\linewidth}
    \color{white} .\hfill.
  \end{minipage}%
  \begin{minipage}[t]{0.34\linewidth}
    \centering
    Insulating Spheres,\\
    Conducting Matrix
  \end{minipage}%
  \begin{minipage}[t][][c]{0.25\linewidth}
    \centering
    Phases match
  \end{minipage}%
  \begin{minipage}[t]{0.34\linewidth}
    \centering
    Conducting Spheres,\\
    Insulating Matrix
  \end{minipage}


  \includegraphics[width=\linewidth]{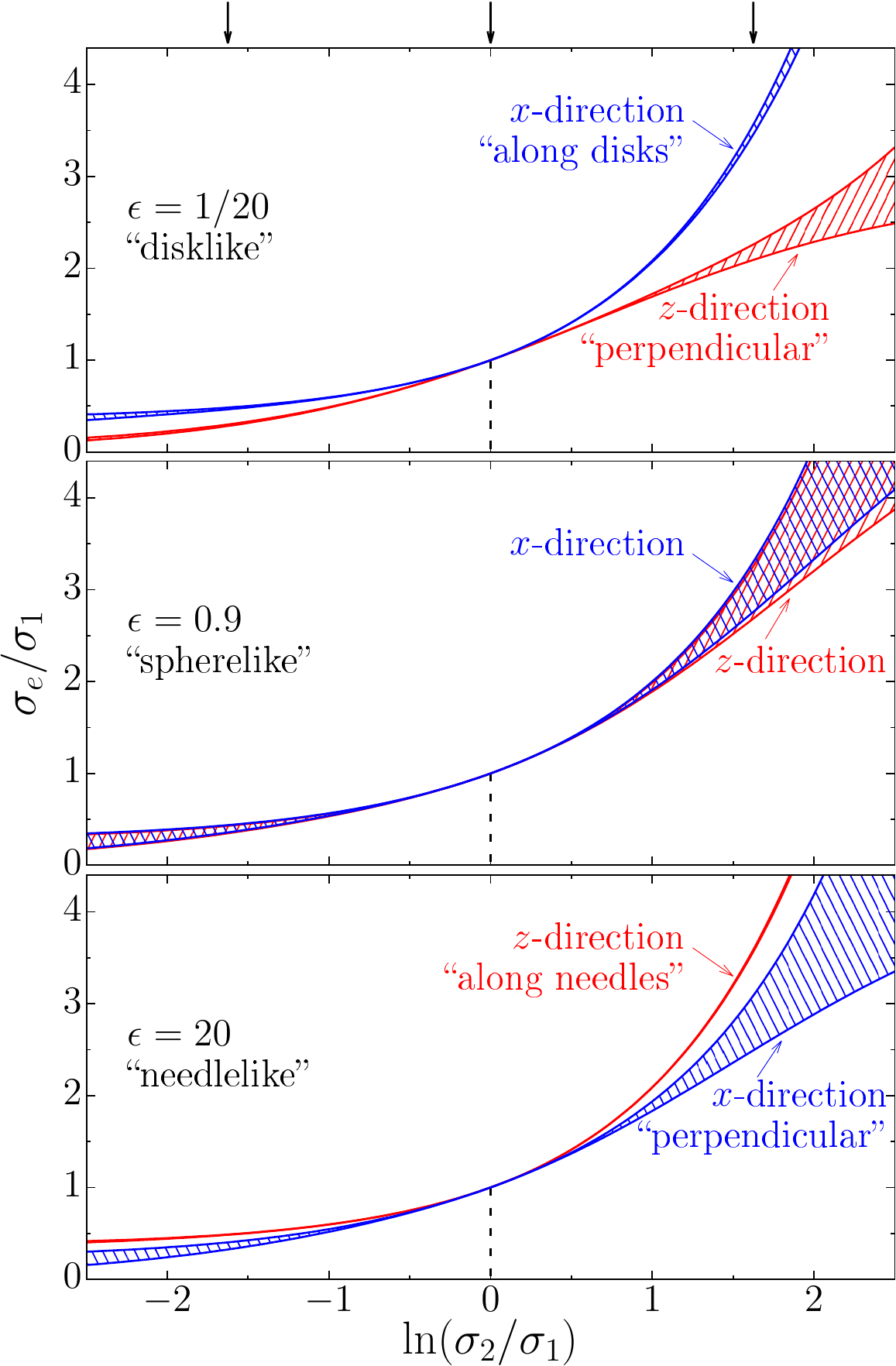}
  \caption{(Color online) Effective conductivity tensor $\mathbf{\sigma_e}$ as a function of the contrast between the conductivities $\sigma_1$ and $\sigma_2$ of the two phases:
The three plots depict the results for different aspect ratios $\epsilon$ from ``disklike'' (top) to ``needlelike'' bottom.
The shaded areas indicate the upper and lower bounds on the diagonal elements of the tensors, distinguishing the $x$- and $z$-directions, where the latter is the direction of the dilation.}
  \label{fig:conductivity}
\end{figure}

In Fig.~\ref{fig:conductivity}, we examine the effective conductivity tensor as a function of the phase contrast $\sigma_2/\sigma_1$ of the two phases spheres (2) and matrix (1).
It plots both the upper and the lower bounds on the diagonal elements $(\sigma_e)_{xx}$ and $(\sigma_e)_{zz}$ of the tensor.
The bounds in $x$- or $z$-direction hardly differ for spheroids that are close to a sphere, see Fig.~\ref{fig:conductivity}~(middle panel),
but the bounds differ distinctly for both the oblate ellipsoids (top panel) and the prolate ellipsoids (bottom panel).
Not only is the material microscopically anisotropic, but the macroscopic physical property becomes anisotropic as well.

Not surprisingly, we find a greater effective conductivity in the orientation of the ``needles'' or in the plane of the ``disks'' independent of which phase is insulating, because both phases become elongated.

Interestingly, the bounds become sharp for the oblate ellipsoids as well as the effective conductivity for prolate ellipsoids in the direction of the symmetry axis~\cite{sen_effective_1989, torquato_conductivity_1990, Torquato2002}.
The explicit rigorous formulas allow for a precise prediction of the effective conductivity.
The sharpness of the bounds can be confirmed analytically.
In the limit $\epsilon \rightarrow 0$, the shape parameter $Q$ vanishes ($Q \rightarrow 0$).
The rigorous upper bounds collapse with the corresponding rigorous lower bounds, see Eqs.~\eqref{eq:boundsigma-A}--\eqref{eq:boundsigma-B}.
The elements of the effective conductivity tensor are exactly given by~\cite{sen_effective_1989, torquato_conductivity_1990}:
\begin{align}
    (\bm{\sigma_e})_{xx} &= \sigma_1\phi_1 + \sigma_2\phi_2, \\
    (\bm{\sigma_e})_{zz} &= \frac{\sigma_1\sigma_2}{\sigma_1\phi_2 + \sigma_2\phi_2},
\end{align}
which are the arithmetic and the harmonic averages of the phase conductivities.
The effective conductivity tensor is the same as for a simple laminate composite, which consists of alternating layers (parallel to the $x$-$y$ plane) of random thickness~\cite{Torquato2002}.

In the limit $\epsilon\rightarrow \infty$, the bounds on $(\bm{\sigma_e})_{xx}$ do not collapse, but converge to the two-dimensional version of the Hashin-Shtrikman bounds for isotropic materials~\cite{Hashin70}, see Eqs.~(21.21) and (21.22) in Ref.~\cite{Torquato2002}.
The bounds on $(\bm{\sigma_e})_{zz}$, however, do collapse to the arithmetic mean:
\begin{align}
    (\bm{\sigma_e})_{zz} &= \sigma_1\phi_1 + \sigma_2\phi_2.
\end{align}

These collapses of the upper and lower bounds imply that
for extreme elongation or compression the effective conductivity (in the preferred orientation) only depends on the volume fraction of the two phases.


\section{Conclusions and Outlook}
\label{sec:Conclusion}

It is the complex geometry that essentially determines the physical properties of heterogeneous materials,
especially for such a remarkable structure as that formed by MRJ sphere packings.
These mechanically stable yet maximally disordered packings are locally disordered, but are globally uniform due to their hyperuniformity property.
In the first two papers of this series, we have studied in detail subtle and distinct geometrical features of the MRJ sphere packings and characterized the structure across length-scales by a score of various characteristics.
Here  we show how these features affect the physical properties of MRJ sphere packings.

We use bounds and approximations based on the structure characteristics that we determined in the second paper.
Thus, we predict flow properties as well as the response to propagating electromagnetic waves and we study the trapping problem in the void phase of MRJ spheres packings.
These predictions are compared both to crystal lattices and to non-hyperuniform systems, overlapping spheres and an equilibrium hard-sphere liquid below the freezing point.

Although rigorous bounds on the effective properties using limited two-point and pore-size information cannot be expected to yield exact predictions of the properties, they allow one
to study the qualitative behavior of different structures without the need of time-consuming simulations.
More precise quantitative predictions can easily be made using ratios of the bounds to those of a system with well-known physical properties.
Moreover, the results are confirmed by employing universal scaling relations.
They provide guidance to carry out more expensive simulation procedures, \eg, of flow profiles using finite-element methods, and experiments
for the design of hyperuniform meta-materials with unique bulk properties, like a transport of electromagnetic waves that is dissipationless simultaneously in all directions.

Stealthy hyperuniformity makes it possible to design isotropic materials with large and complete photonic band gaps~\cite{florescu_designer_2009}.
The MRJ sphere packing is, strictly speaking, not stealthy hyperuniform, but has a linearly decreasing structure factor (with values close to zero for a considerable range and hence nearly stealthy).
It could therefore be used to produce photonic materials with desirable structural-color~\cite{Ballato:00,ADMA:ADMA200903693,PhysRevE.90.062302,wilts_butterfly_2017} or color-sensing~\cite{PhysRevE.89.022721} characteristics.
Additive manufacturing fabrication techniques offer a simple production of samples for experiments with microwaves.

Summarizing our results for flow and diffusion processes in the void phase of the MRJ sphere packings,
the void and interfacial correlation functions allow for rigorous bounds on the fluid permeability as well as on the mean survival time or the closely related trapping constant.
The lower bounds for the trapping constant correspond to the inverse of upper bounds on the mean survival time as well as on the permeability.
We have estimated the involved improper integrals from integrals over the correlation functions of finite simulated samples,
taking advantage of the corresponding analytic explicit expressions derived in the second paper of this series.

We find distinctly stronger bounds on the permeability for the MRJ sphere packings (at $\phi=0.636$) compared to an equilibrium hard-sphere liquid at $\phi=0.478$, which can only in part be explained by the larger packing fraction.
It is reasonable to assume that in the hyperuniform MRJ packings large pores and pore channels are suppressed, resulting in a large trapping constant but small permeability.
This could point to extremal properties of the MRJ packings among isotropic two-phase media.

Extending the analysis of the steady-state trapping problem, we study the mean survival time.
We compute both the interfacial upper bounds and a pore-size lower bound and compare these bounds to predictions from the universal scaling of the mean survival time for particulate models.
As indicated by the bounds, also the universal scaling predicts a strongly reduced mean survival time in the MRJ sphere packings compared to the equilibrium liquid at $\phi=0.478$ but also to overlapping spheres at the same $\phi=0.636$ as the MRJ spheres.
The mean survival time and principle relaxation time in another type of strongly hyperuniform systems (in stealthy disordered systems) has been recently studied in Ref.~\cite{zhang_transport_2016}.

Because the hyperuniform MRJ state is more regular than the equilibrium liquid of hard spheres, we expect a distinctly smaller principal relaxation time in the time-dependent trapping problem.
This would, for example, imply a fast decay of the survival probability of a Brownian particle diffusing in the pore space of the MRJ spheres.

Studying the propagation of electromagnetic waves through disordered media, we find that it strongly depends on subtle structural features.
Especially the imaginary part of the effective dielectric constant strongly depends on the microstructure.

For any hyperuniform material, the imaginary part of the effective dielectric constant vanishes exactly through lowest order in perturbation expansion and third order in $k_q$.
Therefore, the MRJ state is a disordered and isotropic two-phase medium that is nearly dissipationless.
The imaginary part is negligible compared to that of overlapping or equilibrium hard spheres.
When a system is hyperuniform and stealthy, that is, the structure factor vanishes at finite wave vectors, the lowest order in the perturbation expansion has exactly no imaginary part (for all orders in $k_q$).

Finally, the results can easily be generalized to packings of spheroids with a fixed orientation by a linear transformation (which corresponds to a stretching the sphere packings).
The aligned spheroids form not only a microscopically anisotropy structure, but the physical properties also become macroscopically anisotropic.
The mean survival time depends on the aspect-ratio of the particles but remains a scalar quantity.
In contrast to this, the fluid permeability and conductivity have to be represented by tensors that indicate the preferred orientation of the flow or current.
In the limit of infinite compression or elongation, the rigorous two-point bounds on the effective conductivity in the preferred orientation collapse to a function that only depends on the volume fraction of the two phases.

\begin{acknowledgments}
  We thank Steven Atkinson for his simulated samples of MRJ packings and hard-sphere liquids.
  We also thank the German Research Foundation [Deutsche Forschungsgemeinschaft (Germany)] for Grants No. HU1874/3-2 and No. LA965/6-2 awarded as part of the DFG-Forschergruppe FOR 1548 ``Geometry and Physics of Spatial Random Systems.''
  This work was in part supported by the National Science Foundation under Award No. DMR-1714722.
\end{acknowledgments}

\appendix


\section{Integrating correlation functions derived from finite simulation boxes}
\label{sec_integration_S2}

In the second paper of this series, we derived explicit formulas for $S_2(r)$, $F_{sv}(r)$, and $F_{ss}(r)$ for finite packings of hard spheres.
They allow for a fast and accurate calculation of the integrals that are needed for void and interfacial bounds~\cite{RubinsteinTorquato1988, RubinsteinTorquato1989JFM, TorquatoRubinstein1989, Doi1976, Torquato1991Review}.
For finite packings, we can analytically evaluate the corresponding integrals of these correlation functions.
However, the void and interfacial bounds on the permeability $k$ and trapping constant $\gamma$ invoke improper integrals, which we must estimate from the integrals over correlation functions of finite packings.

\subsection{Integral for the void bounds}

First we study the key integral that is needed for the void bounds
\begin{align}
  I(x) & := \int_{0}^{x}r [S_2(r)-\phi^2]\mathrm{d}r,
  \label{eq:def-I}
\end{align}
where $\chi(r):=S_2(r)-\phi^2$ is the autocovariance.
The integral has units length squared.
We have analytically evaluated a lengthy but explicit expression for $I(x)$ for arbitrary values of the upper limit $x$ of the integral.

\begin{figure}[t]
  \centering
  \includegraphics[width=\linewidth]{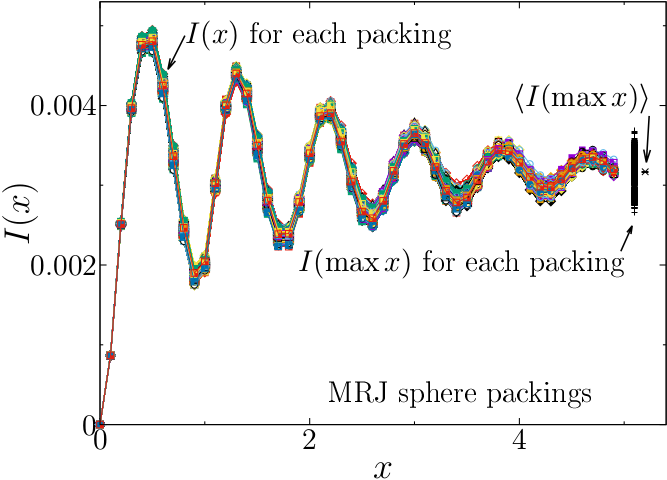}
  \caption{(Color online) The integral $I(x)$ for each finite MRJ sphere packing oscillates but converges to the unbounded integral $\int_{0}^{\infty}r[S_2(r)-\phi^2]\mathrm{d}r$.
    Although the integrals over the finite packings still deviate significantly from this limit, the average over several packings appears to provide a robust estimate.
    However, depending on variations in the cutoff radius, an unknown systematic bias might appear.
  The unit of length is the diameter of a single sphere.}
  \label{fig_Integral_S2}
\end{figure}

While the accessible values of $x$ are limited by the finite sizes of the simulation boxes of the various samples, the void bound uses the improper integral $I:=\lim_{x\rightarrow\infty}I(x)$.
A straightforward approximation is to evaluate $I(x)$ for the maximal possible value of $x$.
This naive approach, however, might induce an unknown systematic bias.
Because of the oscillations in the two-point correlation function, also the integral $I(x)$ oscillates as a function of $x$, see Fig.~\ref{fig_Integral_S2}.

A constant choice of the maximum value of $x$ for different samples would almost surely induce a constant offset.
Because the box sizes vary for the samples produced by the Torquato-Jiao algorithm, also the maximum values of $x$ vary and the average over several samples appears to produce reliable estimates of the improper integral, see Fig.~\ref{fig_Integral_S2}.
However, for a rigorous analysis we must extrapolate the integrals $I(x)$ for each single packing before averaging over all samples.

For a numerically robust extrapolation, we apply the trick of ``$\alpha$-convergence,''
that is, we introduce an additional factor $\exp(-\alpha r)$ to the integrand of Eq.~\eqref{eq:def-I}:
\begin{align}
  I(x,\alpha) & := \int_{0}^{x} \mathrm{e}^{-\alpha r} r (S_2(r)-\phi^2)\mathrm{d}r.
  \label{eq:def-Ialpha}
\end{align}
Note that $I(x)= \lim_{\alpha\rightarrow 0}I(x,\alpha)$ and therefore $I=\lim_{x\rightarrow\infty}I(x)=\lim_{x\rightarrow\infty}\lim_{\alpha\rightarrow 0}I(x,\alpha)$.

\begin{figure}[t]
  \centering
  \includegraphics[width=0.97\linewidth]{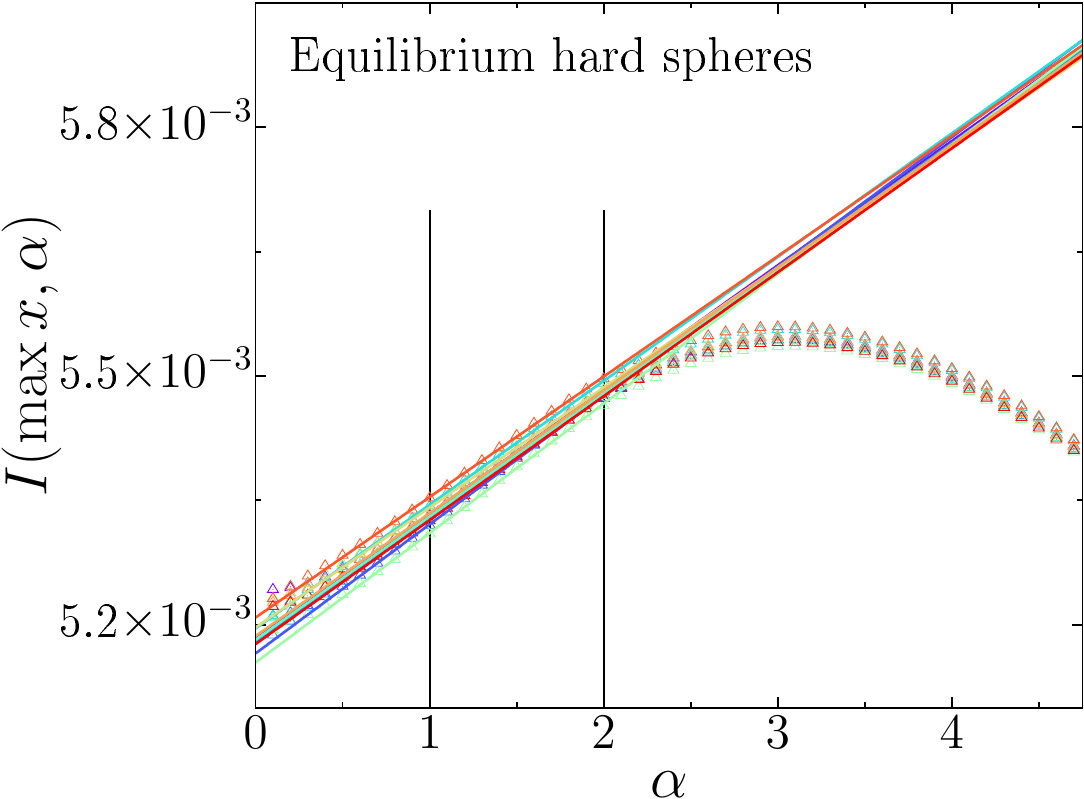}\\
  \includegraphics[width=0.97\linewidth]{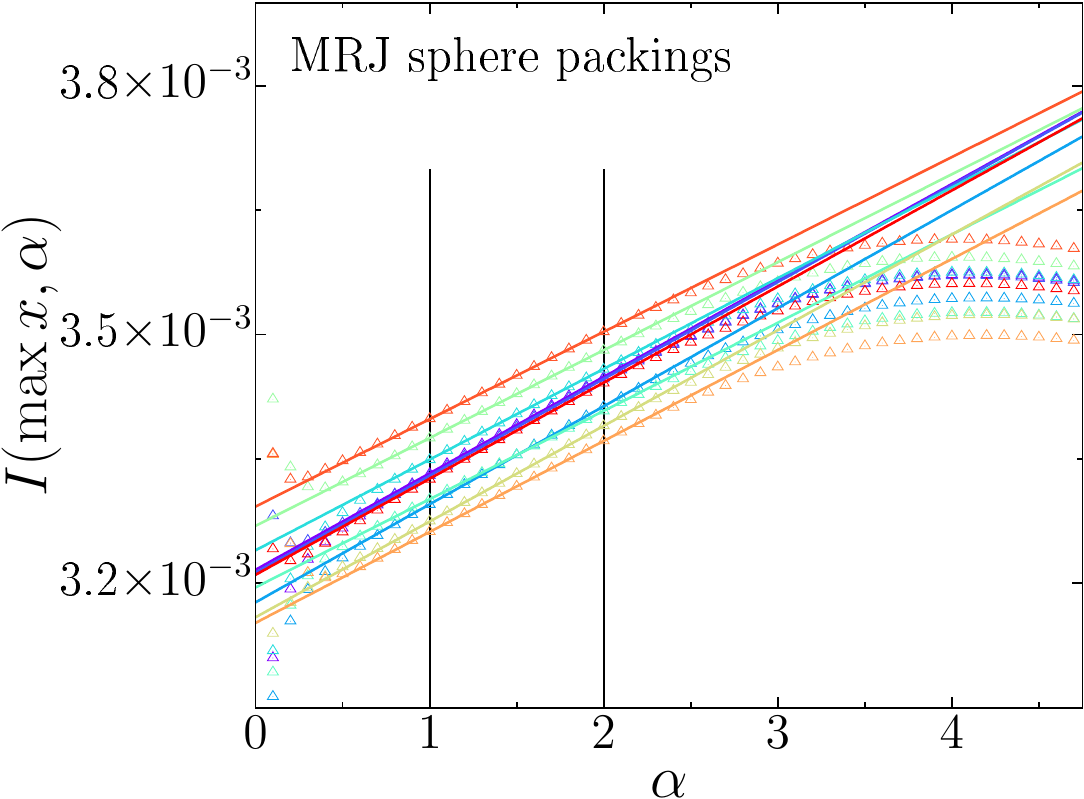}
  \caption{(Color online) Extrapolation of the integral of the two-point correlation function for the void bound, see Eq.~\eqref{eq:def-Ialpha}:
  for $\alpha>1$, the limit $\lim_{x\rightarrow\infty}I(x,\alpha)$ is well approximated by $I(\max x,\alpha)$.
  Thus, $I(\max x,\alpha)$ with $\alpha\in(1,2)$ allows for a robust extrapolation to $I$ (\ie, $\alpha\rightarrow 0$) for both equilibrium (top) and MRJ (bottom) hard spheres.
  Each plot shows the results for ten packings.
  The unit of length is the diameter of a single sphere.}
  \label{fig_Integral_S2_alpha_convergence}
\end{figure}

For non-vanishing values of $\alpha$, the factor $\exp(-\alpha r)$ suppresses the oscillations at large values of $x$,
and therefore $\lim_{x\rightarrow\infty}I(x,\alpha)$ can well be approximated by $I(\max x,\alpha)$, where $\max x$ is the maximum possible value of $x$.
Extrapolating $I(\max x,\alpha)$ for $\alpha\rightarrow 0$ using intermediate values of $\alpha$ thus provides a reliable estimate of $I$ (assuming that the limits can be interchanged).
Figure~\ref{fig_Integral_S2_alpha_convergence} visualizes this extrapolation for single samples of MRJ packings as well as the equilibrium hard-sphere fluid.

We extrapolate the integral for each packing and calculate for each the void bounds from Eqs.~\eqref{eq:void-bound-trapping} and \eqref{eq:void-bound-permeability}.
For the equilibrium hard-sphere liquid, the average over 100 samples, each containing 10000 spheres, yielded a void bound $7.640(4)\gamma_s$ for the trapping constant and $0.13090(6)k_s$ for the fluid permeability.
The normalizing constant $\gamma_s:=12\phi/D^2$ corresponds to the well-known Smoluchowski result for the dilute limit of a spherical trap of diameter $D$.

The average of the bounds over 1015 MRJ packings, each containing 2000 spheres, yielded the void bounds $16.48(1)\gamma_s$ and $0.06068(3)k_s$, respectively.
Despite the small statistical error, systematic errors due to a finite system size at the order of magnitude $\mathcal{O}(0.1\gamma_s)$ [or $\mathcal{O}(0.01k_s)$] can still remain.
Therefore, we have also analyzed 16 MRJ packings with 10000 spheres; the estimated void bounds are $16.76(2)\gamma_s$ and $0.05965(8)k_s$.
These latter values are used in Secs.~\ref{sec:permeability-trapping} and \ref{sec:mean-survival-time}.

The systematic bias in the simple estimate via an average of the integrals evaluated at the maximum value of $x$ turns out to be small
for the MRJ sphere packings analyzed here; the resulting void bounds are $16.78(4)\gamma_s$ for the trapping constant and $0.0598(1)k_s$ for the fluid permeability.
It actually coincides with the estimate based on the extrapolated large MRJ packings.

\subsection{Integrals for the interfacial bound}

\begin{figure}[t]
  \centering
  \includegraphics[width=0.97\linewidth]{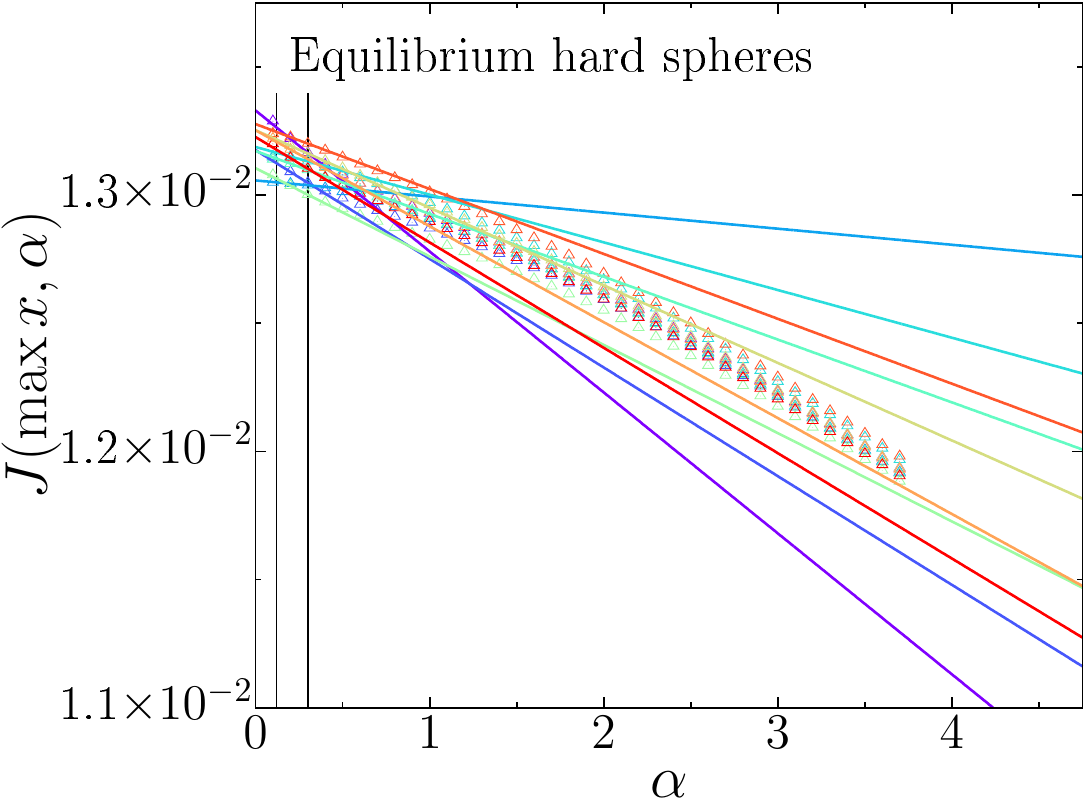}\\
  \includegraphics[width=0.97\linewidth]{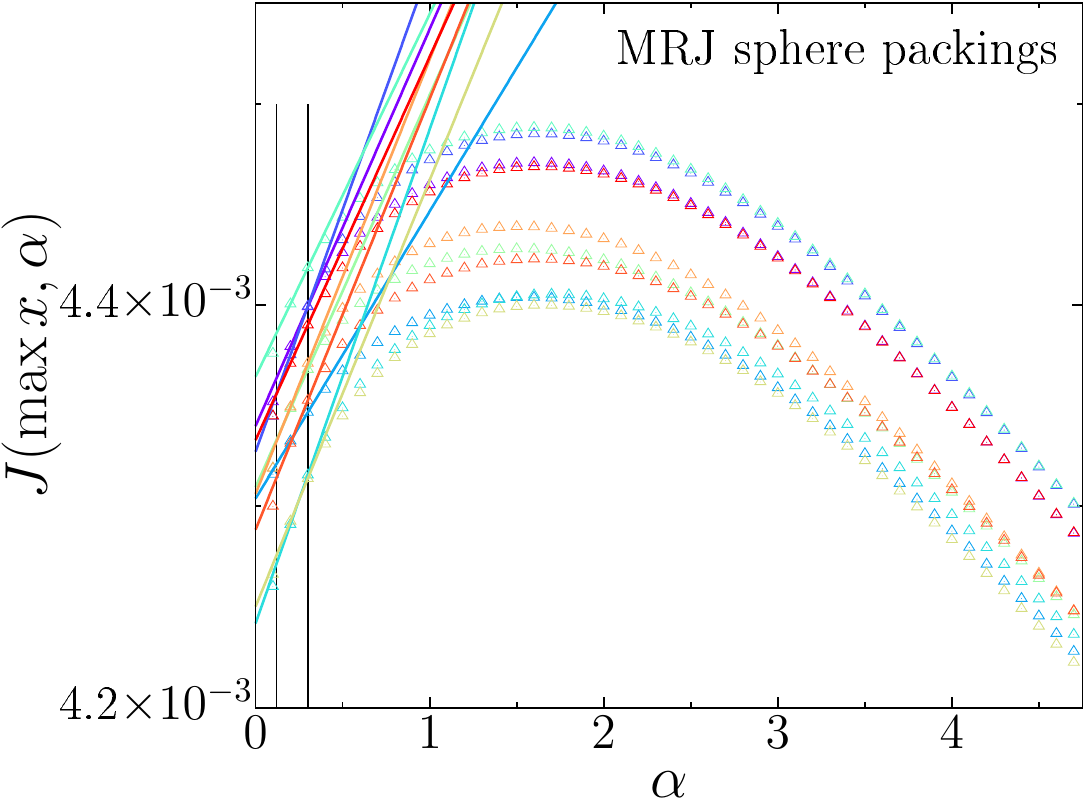}
  \caption{(Color online) Extrapolation of the integral of the correlation functions for the interfacial bound, see Eq.~\eqref{eq:def-Jalpha}:
  for $\alpha>1$, the limit $\lim_{x\rightarrow\infty}J(x,\alpha)$ is well approximated by $J(\max x,\alpha)$.
  Thus, $J(\max x,\alpha)$ with $\alpha\in(0.12,0.30)$ allows for a robust extrapolation to $J$ (\ie, $\alpha\rightarrow 0$) for both equilibrium (top) and MRJ (bottom) hard spheres.
  Each plot shows the results for ten packings, where each contains 10000 spheres.
  The unit of length is the diameter of a single sphere.}
  \label{fig_Integral_J_S2_alpha_convergence}
\end{figure}

For the interfacial bounds we have to estimate the improper integral in Eqs.~\eqref{eq:interfacial-bound-permeability} and \eqref{eq:tau-upper-bound}.
The same problems and procedures apply to this integral as for the integral discussed in the preceding section.
Again we have derived explicit expressions for both the proper integrals and the versions for ``$\alpha$-convergence'' with the additional factor $\exp(-\alpha r)$:
\begin{align}
  \begin{aligned}
    J(x,\alpha) := \int_{0}^{x} \mathrm{e}^{-\alpha r} r \big[&\frac{(1-\phi)^2}{s^2}F_{ss}(r)\big.\\
    & \big.-\frac{2(1-\phi)}{s}F_{sv}(r)+F_{vv}(r)\big]\mathrm{d}r,
  \end{aligned}
  \label{eq:def-Jalpha}
\end{align}
and with a fit in the range $0.12<\alpha<0.30$.
Figure~\ref{fig_Integral_J_S2_alpha_convergence} shows examples of the extrapolation for both MRJ and equilibrium hard-sphere liquids.
For each packing, we extrapolate the integral (analogously to the preceding section) and calculate the interfacial bounds using Eqs.~\eqref{eq:interfacial-bound-permeability} and \eqref{eq:tau-upper-bound}.

The average over the 100 samples of an equilibrium hard-sphere liquid provided the interfacial bounds $13.06(1)\gamma_s$ for the trapping constant and $0.07661(8)k_s$ for the fluid permeability.
From the MRJ packings with 2000 spheres, we derived the interfacial bounds $29.82(2)\gamma_s$ and $0.03355(2)k_s$, respectively.
However, in Secs.~\ref{sec:permeability-trapping} and \ref{sec:mean-survival-time}, we use the bounds based on the MRJ packings with 10000 spheres: $30.42(7)\gamma_s$ and $0.03287(7)k_s$.

\section{Pore sizes and bounds for sphere packings on Bravais lattices}
\label{sec_lattices}

\begin{figure}[b]
  \centering
  \subfigure[][]{\includegraphics[width=0.32\linewidth]{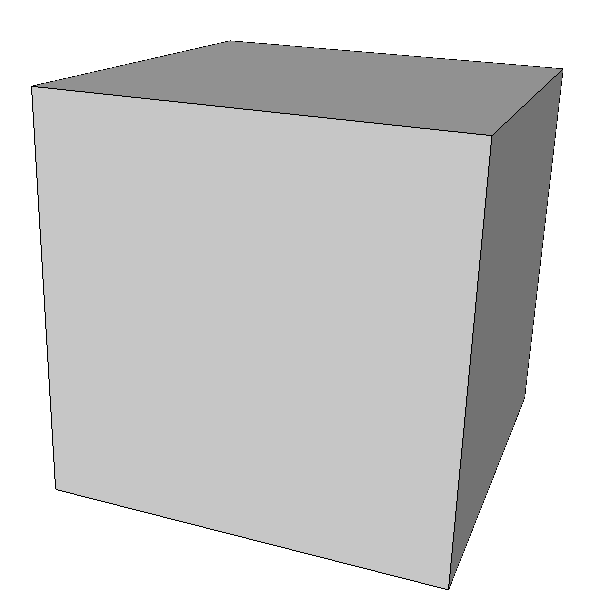}}
  \subfigure[][]{\includegraphics[width=0.32\linewidth]{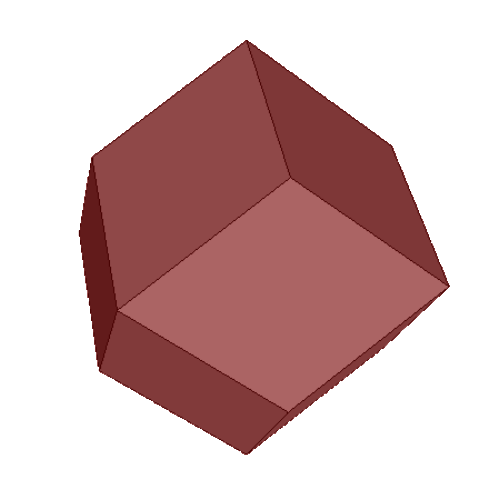}}
  \subfigure[][]{\includegraphics[width=0.32\linewidth]{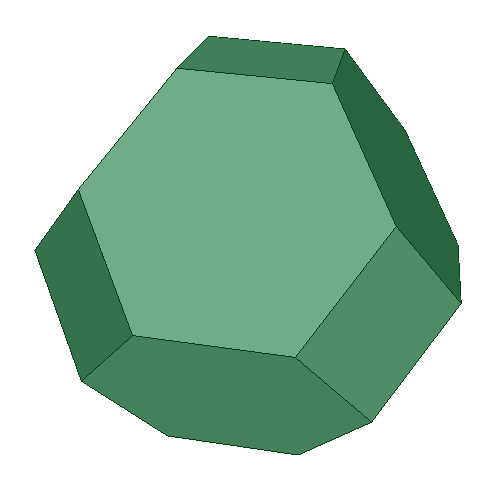}}
  \caption{(Color online) Wigner-Seitz (Voronoi) cells of three Bravais lattices: (a) SC, (b) FCC, and (c) BCC.}
  \label{fig:ws-cells}
\end{figure}

Figure~\ref{fig:ws-cells} shows Wigner-Seitz cells $W$ of three Bravais lattices.
The volume of the Wigner-Seitz cell is, in the following, denoted by $v_W$.
We consider a packing of hard, monodisperse spheres with diameter $D=2R$, the centers of which coincide with the lattice points.
The volume of such a sphere $B$ is $v_B = \pi D^3/6$.
The unit of length is defined by the maximal diameter of spheres (without overlap), which corresponds to the maximum packing fraction.
Here the lattice spacing is fixed; lower packing fractions $\phi=v_B/v_W$ correspond to smaller sphere diameters.

\subsection{Moments of pore sizes}
By definition of the Wigner-Seitz cell (as a Voronoi cell), the lattice point that is closest to any of the points in its interior is the origin.
Because in a Bravais lattice all Voronoi cells are congruent to each other and (because a point in a Poisson point process is almost surely inside some Voronoi cell),
all moments of the pore-size $\delta$ distribution of a lattice can be evaluated by corresponding integrals over (the interior of) the Wigner-Seitz cell only.
\begin{align}
  \langle \delta^n \rangle &= \frac{ \int_{W\backslash B} (|x|-R)^n d^3x}{v_W-v_B} \\
                           &= \frac{ \int_{W} (|x|-R)^n d^3x - \int_{B} (|x|-R)^n d^3x}{v_W-v_B} \\
			   &= \frac{\int_{W} (|x|-R)^n d^3x - 4\pi \int_{0}^{R} r^2(r-R)^n dr}{v_W-v_B} \\
			   &= \frac{ \int_{W} (|x|-R)^n d^3x }{v_W-v_B} \nonumber\\
			   & \quad + \frac{4\pi R^{n+3}}{v_W-v_B} \left[\frac{(2n+4)(-1)^{n+1} }{n^2+4n+3} + \frac{2(-1)^{n}}{n+2}   \right] 
  \label{eq:moments-as-integrals}
\end{align}
For the mean pore size ($n=1$), we find
\begin{align}
  \langle \delta \rangle &= \frac{\int_{W} |x| d^3x - \frac{1}{2} v_W D + \frac{\pi}{48} D^{4} }{v_W-v_B}
  \label{eq:mean}
\end{align}
For the second moment $n=2$, we find
\begin{align}
  \langle \delta^2 \rangle &= \frac{\int_{W} |x|^2 d^3x - D \int_{W} |x| d^3x + \frac{1}{4} v_W D^2 - \frac{\pi}{240} D^{5}}{v_W-v_B}
  \label{eq:scnd}
\end{align}
The first integral in the last equation is equal to the trace of a Minkowski tensor\cite{schroder-turk_minkowski_2011,klatt_mean-intercept_2017}:
\begin{align}
  \text{tr}W_0^{2,0}(W) = \int_{W} |x|^2 d^3x.
  \label{eq:MT}
\end{align}
Note that the integrals in Eqs.~\eqref{eq:mean} and \eqref{eq:scnd}
\begin{align}
  C_1(W) &:= \int_{W} |x| d^3x, \\
  C_2(W) &:= \int_{W} |x|^2 d^3x .
\end{align}
are independent of the sphere diameters (and thus also of the packing fraction).
Once these constants of a lattice (as well as the constant volume $v_W$) are determined, the first and second moments of the pore-size distribution are given as explicit functions of the packing fraction:
\begin{align}
  \label{eq:1}
  \frac{\langle \delta \rangle(\phi)}{D} &=\frac{1}{1-\phi}\left[ \left(\frac{\pi}{6\phi}\right)^{1/3}\frac{C_1(W)}{v_W^{4/3}}- \frac{1}{2} + \frac{\phi}{8} \right] \\
  \frac{\langle \delta^2 \rangle(\phi)}{D^2} &= \frac{1}{1-\phi} \left[ \left(\frac{\pi}{6\phi}\right)^{2/3}\frac{C_2(W)}{v_W^{5/3}} - \left(\frac{\pi}{6\phi}\right)^{1/3}\frac{C_1(W)}{v_W^{4/3}} \right. \nonumber\\
      & \qquad\qquad\quad \left. + \frac{1}{4} -  \frac{\phi}{40} \right] 
  \label{eq:2}
\end{align}
using $D^{-3} = \pi/(6v_W\phi)$.
Note that ${C_2(W)}/{v_W^{5/3}}$ is the scaled dimensionless lattice quantizer error~\citep{QuantiserProblem}.
The moments of the pore size distributions could be determined via a numerical integration of the void exclusion probability, see Refs.~\citep{QuantiserProblem,KlattTorquato2016}.
However, here the quantizer error and the integral for the first moment are expressed as surface integrals, which can then be easily evaluated for (three examples of) cubic lattices.

\subsection{Surface integrals}
The crucial step in an analytic calculation of the first and second moments of the pore-size distribution is to determine the integrals $C_1$ and $C_2$.
Using the divergence theorem, these volume integrals can be rewritten as surface integrals.
The first integral yields
\begin{align}
  C_1(W) = \int_{W} \frac{1}{4}\text{div}|x|x d^3x = \frac{1}{4} \oint_{\partial W} |x|x\cdot n dS,
\end{align}
because
\begin{align}
  \text{div}|x|x = 3|x| + x\cdot\nabla|x| = 3|x| + \frac{x\cdot x}{|x|} = 4|x|.
\end{align}
The second integral results in
\begin{align}
  C_2(W) = \int_{W} \frac{1}{5}\text{div}|x|^2x d^3x = \frac{1}{5} \oint_{\partial W} |x|^2x\cdot n dS,
\end{align}
because
\begin{align}
  \text{div}|x|^2x = 3|x|^2 + x\cdot\nabla|x|^2 = 5|x|^2.
\end{align}
Because the Wigner-Seitz cells are polytopes, the integrals can be rewritten as sums of integrals over the two-dimensional facets
\begin{align}
  C_1(W) &= \frac{1}{4} \sum_{F\in \mathcal{F}_2(W)} n \cdot \int_{F}|x|x \mathcal{H}^2(dx), \\
  C_2(W) &= \frac{1}{5} \sum_{F\in \mathcal{F}_2(W)} n \cdot \int_{F}|x|^2x \mathcal{H}^2(dx),
\end{align}
where $\mathcal{H}^2$ denotes the two-dimensional Hausdorff measure.
The calculation for a regular polytope can be further simplified, because each term in the sum is a scalar quantity. 
Faces that can be mapped onto each other by a rotation will yield the same contributions.

\subsection{Three cubic lattices}
For the \textit{simple cubic} (SC) lattice with a cube as the Wigner-Seitz cell, the integrals are given by
\begin{align}
  C_1(W_{\text{SC}}) &= \frac{3D_{\max}}{4} \iint_{-D_{\max}/2}^{D_{\max}/2} \sqrt{\frac{D_{\max}^2}{4}+y^2+z^2} dydz, \\
  C_2(W_{\text{SC}}) &= \frac{3D_{\max}}{5} \iint_{-D_{\max}/2}^{D_{\max}/2}(\frac{D_{\max}^2}{4}+y^2+z^2) dydz,
\end{align}
where $D_{\max}$ is the maximal diameter, which is here set to one by the choice of units.

For the {face-centered cubic} (FCC) lattice, the Wigner-Seitz cell is a rhombic dodecahedron with 12 identical faces.
Its edge length is
  $a_{\text{FCC}}:=\sqrt{{3}/{8}}\,{D_{\max}}$,
and its volume is
  $v_{\text{FCC}}:={D_{\max}^3}/{\sqrt{2}}$.
The corners of one of its faces are given by
$c^{(rd)}_1     :=\frac{\sqrt{2}D_{\max}}{4}(2,0,0)^t$,
$c^{(rd)}_2     :=\frac{\sqrt{2}D_{\max}}{4}(0,0,2)^t$,
$c^{(rd)}_{3,4} :=\frac{\sqrt{2}D_{\max}}{4}(1,\pm 1,1)^t$.
The normal vector is $n=\frac{\sqrt{2}}{2}(1,0,1)^t$.
Two difference vectors of the corners are $\Delta_1:=c^{(rd)}_3 - c^{(rd)}_1$ and  $\Delta_2:=c^{(rd)}_4 - c^{(rd)}_1$.
A parametrization of the face by two parameters $s,t\in[0,1]$ is $c^{(rd)}_1 + s \Delta_1 + t \Delta_2$.
For an integration based on this parametrization, we need the Jacobian determinant $J=\sqrt{2}D_{\max}^2/4$.
Explicit expressions for the integrals $C_1$ and $C_2$ are then given by
\begin{widetext}
\begin{align}
  C_1(W_{\text{FCC}}) &= \frac{12}{4\sqrt{2}}\iint_0^1 \frac{2D_{\max}^2}{16}(2-t-s+t+s) \cdot \sqrt{(2-t-s)^2+(t-s)^2+(t+s)^2}\;J\;dsdt\\
                      &= \frac{3D_{\max}^4}{16}\iint_0^1 \sqrt{(2-t-s)^2+(t-s)^2+(t+s)^2}\;dsdt, \\
  C_2(W_{\text{FCC}}) &= \frac{3\sqrt{2}D_{\max}^4}{80}\iint_0^1 ((2-t-s)^2+(t-s)^2+(t+s)^2)dsdt.
\end{align}
\end{widetext}

For the {body-centered cubic} (BCC) lattice, the Wigner-Seitz cell is a trucated octahedron, which is formed by 6 squares and 8 hexagons.
Its edge length is
  $a_{\text{BCC}}:={D_{\max}}/{\sqrt{6}}$,
and its volume is
  $v_{\text{BCC}}:={4D_{\max}^3}/{(3\sqrt{3})}$.
The sphere does not touch the squares, which are at a distance $D_{\max}/\sqrt{3}$ from the origin.
The integral over these squares is similar to that for the simple cubic lattice.
The contributions from the squares are given by:
\begin{widetext}
\begin{align}
  C_{1,\text{squares}}(W_{\text{BCC}}) &= \frac{\sqrt{3}D_{\max}}{2} \iint_{-\frac{D_{\max}}{2\sqrt{6}}}^{\frac{D_{\max}}{2\sqrt{6}}} \sqrt{\frac{D_{\max}^2}{3}+y^2+z^2}\; dydz, \\
  C_{2,\text{squares}}(W_{\text{BCC}}) &= \frac{2\sqrt{3}D_{\max}}{5} \iint_{-\frac{D_{\max}}{2\sqrt{6}}}^{\frac{D_{\max}}{2\sqrt{6}}} \left(\frac{D_{\max}^2}{3}+y^2+z^2\right) dydz.
\end{align}
\end{widetext}
The sphere touches each hexagon. Their distance to the origin is therefore $D_{\max}/2$.
For convenience, the integral over each hexagon can be split in an integral over a rectangle (with side-lengths $a_m$ and $\sqrt{3}a_m$)
and over two (symmetric) triangles.
The contributions from the first integrals are (again similar to the SC lattice and) given by:
\begin{widetext}
\begin{align}
  C_{1,\text{rectangles}}(W_{\text{BCC}}) &= D_{\max} \int_{-\frac{D_{\max}}{2\sqrt{6}}}^{\frac{D_{\max}}{2\sqrt{6}}}\int_{-\frac{D_{\max}}{2\sqrt{2}}}^{\frac{D_{\max}}{2\sqrt{2}}} \sqrt{\frac{D_{\max}^2}{4}+y^2+z^2} \; dydz, \\
  C_{2,\text{rectangles}}(W_{\text{BCC}}) &= \frac{4D_{\max}}{5} \int_{-\frac{D_{\max}}{2\sqrt{6}}}^{\frac{D_{\max}}{2\sqrt{6}}}\int_{-\frac{D_{\max}}{2\sqrt{2}}}^{\frac{D_{\max}}{2\sqrt{2}}} \left(\frac{D_{\max}^2}{4}+y^2+z^2\right) dydz.
\end{align}
\end{widetext}
To calculate the integral over the triangles, consider one representative where the vertices are $(\frac{D_{\max}}{2}, \pm\frac{D_{\max}}{2\sqrt{2}}, \frac{D_{\max}}{2\sqrt{6}})$ and $(\frac{D_{\max}}{2}, 0, \frac{D_{\max}}{\sqrt{6}})$.
The bounds of the integral of the $y$-component depend on the $z$-coordinate: $f(z)=D_{\max}/\sqrt{2}-\sqrt{3}z$.
The final contributions of the integrals are then given by:
\begin{widetext}
\begin{align}
  C_{1,\text{triangles}}(W_{\text{BCC}}) &= 2D_{\max} \int_{\frac{D_{\max}}{2\sqrt{6}}}^{\frac{D_{\max}}{\sqrt{6}}}\int_{-f(z)}^{f(z)} \sqrt{\frac{D_{\max}^2}{4}+y^2+z^2}\; dydz, \\
  C_{2,\text{triangles}}(W_{\text{BCC}}) &= \frac{8D_{\max}}{5} \int_{\frac{D_{\max}}{2\sqrt{6}}}^{\frac{D_{\max}}{\sqrt{6}}}\int_{-f(z)}^{f(z)} \left(\frac{D_{\max}^2}{4}+y^2+z^2\right) dydz.
\end{align}
\end{widetext}
The final constants $C_1(W_{\text{BCC}})$ and $C_2(W_{\text{BCC}})$ are then given by the sum of these contributions.

The final integrals for SC, FCC, and BCC lattice packings were evaluated via numerical integration using \textsc{Maple} 18~\cite{maple}.
\begin{align}
\label{eq:analytic-results-first}
  C_1(W_{\text{SC}})  &\approx 0.480295978227526,\\
  C_2(W_{\text{SC}})  &= \frac{1}{4} = 0.25,\\
  C_1(W_{\text{FCC}}) &\approx 0.295745942288160,\\
  C_2(W_{\text{FCC}}) &= \frac{3\sqrt{2}}{32} \approx 0.132582521472478,\\
  C_1(W_{\text{BCC}}) &\approx 0.330947379746666,\\
  C_2(W_{\text{BCC}}) &= \frac{19\sqrt{3}}{216} \approx 0.152356321036154.
  \label{eq:analytic-results-last}
\end{align}

The resulting moments of the pore size (evaluated with 15 digits numerical precision) are for the SC, FCC, and BCC lattice packings at their corresponding maximum packing fractions given by:
\begin{align}
\langle \delta \rangle_{\min}(\text{SC})    &\approx 0.0960237355283125,\\
\langle \delta^2 \rangle_{\min}(\text{SC})  &\approx 0.0138833656248951,\\
\langle \delta \rangle_{\min}(\text{FCC})   &\approx 0.0416461321138215,\\
\langle \delta^2 \rangle_{\min}(\text{FCC}) &\approx 0.00285167445050149,\\
\langle \delta \rangle_{\min}(\text{BCC})   &\approx 0.0466976980202309,\\
\langle \delta^2 \rangle_{\min}(\text{BCC}) &\approx 0.00312370712463761.
\end{align}
To mutually check the simulations and the analytic calculations, these numerical integrals can be compared to simulation results for these lattice-packings based on $10^7$ sampling points:
\begin{align}
\langle \delta \rangle^{\text{sim}}_{\min}(\text{SC})    &\approx 0.09604(2),\\
\langle \delta^2 \rangle^{\text{sim}}_{\min}(\text{SC})  &\approx 0.013887(5),\\
\langle \delta \rangle^{\text{sim}}_{\min}(\text{FCC})   &\approx 0.04163(1),\\
\langle \delta^2 \rangle^{\text{sim}}_{\min}(\text{FCC}) &\approx 0.002850(1),\\
\langle \delta \rangle^{\text{sim}}_{\min}(\text{BCC})   &\approx 0.04668(1),\\
\langle \delta^2 \rangle^{\text{sim}}_{\min}(\text{BCC}) &\approx 0.003122(1).
  \label{eq:simulation-results}
\end{align}
They are in excellent agreement with each other.

\begin{figure}[t]
  \centering
  \subfigure[][]{\includegraphics[width=\linewidth]{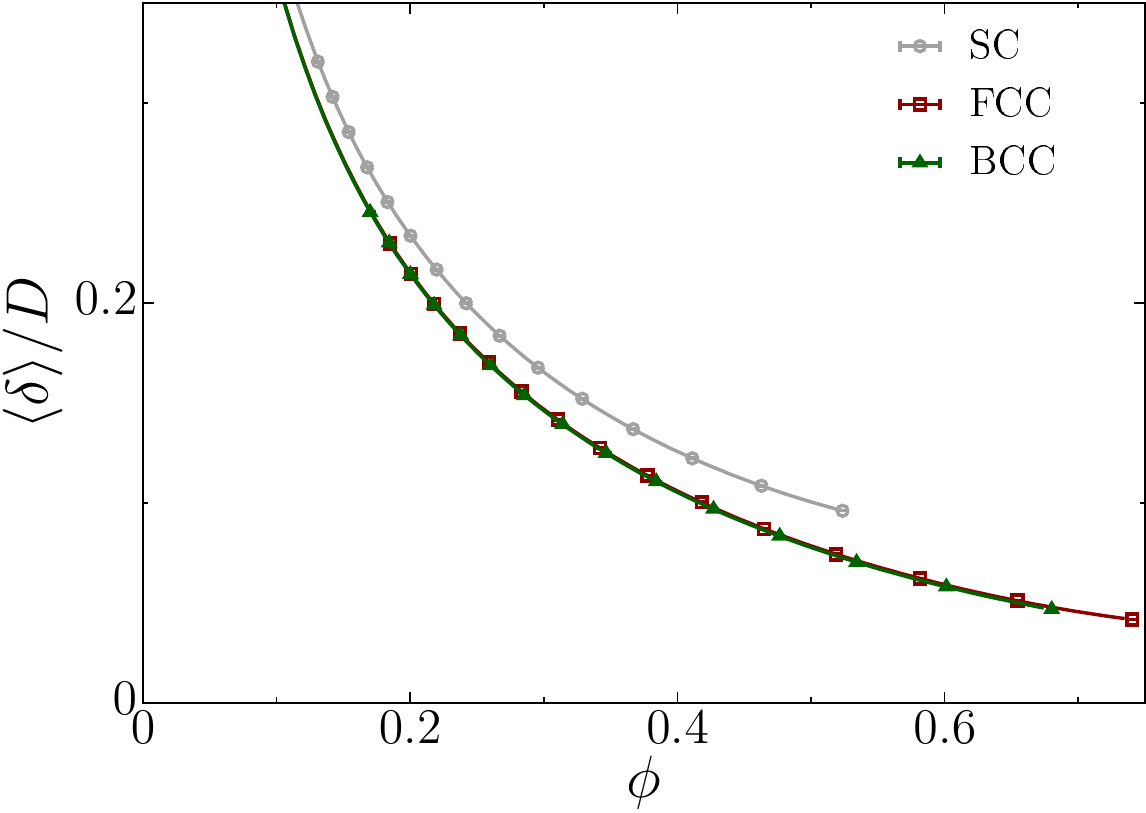}}
  \subfigure[][]{\includegraphics[width=\linewidth]{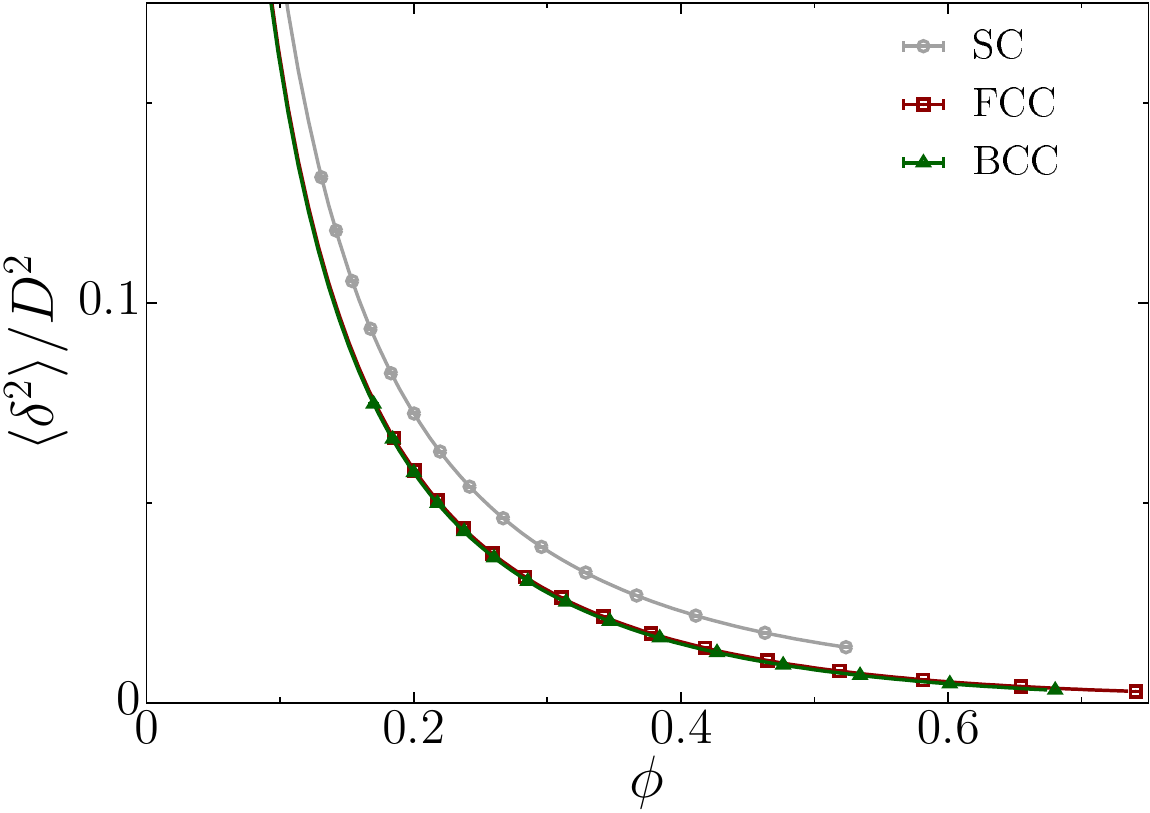}}
  \caption{(Color online) For the three Bravais lattices FCC, BCC, and SC, the analytic curves and simulation results for the mean pore size $\langle \delta \rangle$ and the second moment $\langle \delta^2 \rangle$.}
  \label{fig:pore-sizes}
\end{figure}

Using Eqs.~\eqref{eq:analytic-results-first}--\eqref{eq:analytic-results-last}, \eqref{eq:1}, and \eqref{eq:2},
the pore sizes of lattice packings at arbitrary packing fractions can easily be calculated.
Figure~\ref{fig:pore-sizes} compares the analytic results to simulations.
They are in excellent agreement.

%

\end{document}